\documentclass[12pt,preprint]{aastex}
\def\etal{{et al.}}

\def\ex{{\it EXOSAT}}

\def\xte{{\it RXTE}}

\def\chisq{{$\chi^{2}$}}

\def\Msun{\hbox{$\rm\thinspace M_{\odot}$}}
\begin{document}

\title {Temporal characteristics of the X-ray emission of NGC 7469}

\author {K. Nandra\altaffilmark{1, 2} and
I. E. Papadakis\altaffilmark{3}
}

\altaffiltext{1}{Laboratory for High Energy Astrophysics, Code 662, 
	NASA/Goddard Space Flight Center,
  	Greenbelt, MD 20771}
\altaffiltext{2}{Universities Space Research Association}
\altaffiltext{3}{Department of Physics, University
of Crete, 71003, Heraklion, Greece}

\slugcomment{Accepted for publication in
{\em The Astrophysical Journal}}

\begin{abstract}

We present a study of the time variability of NGC 7469, based on a
$\sim 30$~d \xte/PCA observation. Variability is seen across the X-ray
band, and spectral changes are observed. The softness ratio is
correlated with the ultraviolet flux, but appears to show more rapid
variability. The RMS variability parameter calculated on $\sim 1$d
time scales is significantly variable, and sharp increases are seen
close to the periods where the spectrum is hardest. Cross correlation
of 2-4 keV, 4-10 keV and 10-15 keV light curves show a peak at zero
delay, but the CCFs are skewed towards positive lags (i.e. the soft
leading the hard) with an apparent time lag of about 0.5d. The power
spectral density (PSD) function in the 2-10 keV band shows no clear
features - for example periodicities or breaks - but a power law is a
rather poor fit, particularly to the high frequency spectrum. The
normalized PSD in the soft X-ray band shows a larger amplitude of
variability on long time scales but the hard X-ray PSD is flatter, and
shows more power on time scales less than about 1d. Our data broadly support
the idea that the X-rays are produced by Compton upscattering of
lower-energy seed photons, as had been previously concluded. They are
difficult to reconcile with models in which the sole source of
variability is the seed photons, and more likely suggest a variability
process intrinsic to the X-ray corona.  Our interpretation for these
results is that the low-frequency variability in NGC 7469 does arise
via variations in the Compton seed photons, but that the high
frequency variability arises from the coronal heating mechanism. Both
induce spectral variations. A patchy corona is implied, and one
interpretation consistent with the data is that hotter blobs exist
closer to the central black hole, responsible for the most rapid
variations in the hard X-rays.

\end{abstract}


\section{INTRODUCTION}
\label{Sec:Introduction}

X-ray emission is observed ubiquitously in active galactic nuclei
(AGN). The spectral properties of this emission have been reasonably
well characterized. Once ``secondary'' features due to absorption,
emission and scattering in surrounding material are accounted for, it
appears that the underlying spectrum has an approximate power-law form
over a wide range of energies. The photon spectral index clusters
around a typical value of $\Gamma=2.0$, but with a range of values
(e.g. Nandra \& Pounds 1994).  Many sources show a cutoff at higher
energies, which can be modeled as an exponential with e-folding
energy of several $100$~keV (e.g. Madejski et al. 1995; Gondek et
al. 1996).  The power-law nature of the X-ray spectrum leads to the
suggestion that it is produced by Compton upscattering of softer
photons in a hot ``corona'' (e.g. Shapiro, Lightman \& Eardley 1976;
Sunyaev \& Titarchuk 1980). Such models fit the general
characteristics of X-ray spectrum extremely well (Haardt \& Maraschi
1991, 1993), particularly if the corona consists of discrete blobs
(Haardt, Maraschi \& Ghisellini 1994; Stern et al. 1995) and the
particle distribution has a thermal form (Zdziarski et al. 1994). The
coronal heating mechanism is not known, but some possibilities are
that a portion of the accretion flow is intrinsically hot
(e.g. Shapiro, Lightman \& Eardley 1976; Narayan \& Yi 1994) or that
magnetic reconnection creates hot flaring regions above an accretion
disk (e.g. Nayakshin \& Melia 1997; Poutanen \& Fabian 1999).

The characteristics of the temporal variations of the X-ray emission
can also strongly constrain the emission mechanism.  EXOSAT
demonstrated rapid variability in a number of AGN (e.g. Lawrence et
al. 1985; McHardy \& Czerny 1987) and allowed the first definition of
their power spectral density (PSD) function. Lawrence \& Papadakis
(1993) and Green, McHardy \& Lehto (1993) performed systematic
analyses of the EXOSAT data. Both sets of workers generally found a
featureless PSD with a steep ``red noise'' power-law characterizing
the variations. For an assumed form for the power of $P(f)=A
f^{-\alpha}$, where A is the normalization, all objects were
consistent with a single slope of $\alpha=1.5$. This steep slope
indicates that there must be a turnover (that will henceforth be
termed the ``knee'') at low frequencies, or the integral power would
be infinite. Early attempts at finding this knee (e.g. Papadakis \&
McHardy 1995) have now been improved upon using RXTE data, and there
are now several convincing reports (e.g.  McHardy et al. 1998; Edelson
\& Nandra 1999; Chiang et al. 2000). The knee frequencies indicate
characteristic time scales of orders days-months. There has also been
the report in at least one case of a high frequency break, with the
PSD of Seyfert MCG-6-30-15 showing a further steepening above a
frequency of $\sim 10^{-3}-10^{-4}$~Hz (Nowak \& Chiang 2000).  The
general form of the PSD immediately brings to mind that of Cyg X-1
(e.g. Belloni \& Hasinger 1990), which exhibits ``white noise''
($\alpha=0$) variability below the knee, steepening to $\alpha=1$ and
then to $\alpha=2$ above the high frequency break.

The X-ray variability behavior of AGN just described has never been
convincingly of consistently explained. Typical variations are much
faster than the expected variability time scales in the accretion
disk, such as the viscous, thermal or sound-crossing times
(e.g. Molendi, Maraschi \& Stella 1992). The relationship between the
accretion disk and the X-ray source is not well established, however,
so these time scales may not be relevant.  In the Compton upscattering
model, variations in the ``seed'' source (often assumed to be the
disk) would be translated directly into the up-scattered emission, and
changes in the physical characteristics of the coronal plasma would
also induce variability. The variations may therefore have more to do
with the process by which the Comptonizing plasma is
heated/accelerated than the radiative emission mechanism.

One suggestion as to the variability mechanism in AGN, if not specifically
the origin, is the ``hot spot'' model (Abramowicz et al. 1991; Wiita et
al. 1991). This accounts for the variability by invoking active
regions on the accretion disk whose variations are amplified by
relativistic effects.  This is able to account both for the form of
the PSD and the anticorrelation with luminosity by orientation effects
(Bao \& Abramowicz 1996). Models of X-ray variability in Galactic
Black Hole Candidates and neutron star binaries are somewhat better
developed, although they tend to be tailored to the particular
observational properties of those sources. As much less is known about
the X-ray variability of AGN, it is unclear how relevant such models
are. As mentioned above, however, the X-ray variability can arise from
the seed photons for Comptonization and therefore may be intrinsic to
the source of those photons, most often supposed to be the accretion
disk (Malkan \& Sargent 1982).  Instabilities in the disk could
produce variability which would then be translated into the
Comptonized emission. If the inner part of the flow becomes hot enough
to emit X-rays (e.g. Narayan \& Yi 1994) then instabilities in the
flow might produce variability more directly. In magnetic reconnection
models the heating of the corona is expected to be localized and
episodic and would be expected to produce variability naturally (e.g
Poutanen \& Fabian 1999). These processes would be expected to produce
different variability characteristics in detail, they can be tested
and motivation for development provided with precise observations.

Here we present a detailed analysis of the X-ray variability
characteristics of NGC 7469, based on a long ($\sim 30$d),
densely-sampled RXTE observation. A preliminary light curve and PSD
have already been presented by Nandra et al. (1998; hereafter
N98). Data were also obtained simultaneously with IUE (Wanders et
al. 1997).  N98 showed that the 2-10 keV X-ray flux and the IUE
1315\AA\ flux were not well correlated at zero lag, and furthermore
that the the X-rays showed very rapid variations which were not
present in the UV - the supposed seed photons for
Comptonization. Despite this, Nandra et al. (2000; hereafter N2K)
presented strong evidence that the X-rays were indeed produced by
thermal Comptonization, by demonstrating a strong correlation between
the UV and the X-ray spectral index. Here, we investigate what the
temporal variations of the X-rays alone have to tell us about the
emission mechanisms in NGC 7469.

\section{OBSERVATIONS}

The \xte\ observation began on 1996 June 10 and lasted for a period of
approximately 30 days. We restrict ourselves here to analysis of data
from the proportional counter array (PCA). Full details of the PCA
data analysis are given in N2K. Our analysis differs only very
slightly, in that we applied the very latest screening criteria and
background models (faintl7 version 1999-Aug-24 and faint240 version
1999-Sep-09). We discuss the potential effects of errors in the
background model in an Appendix. We analyzed the standard-2 mode data,
which have a minimum time resolution of 16s. The data do not cover the
whole period of the observation uniformly, however. The observation
strategy in the NGC 7469 campaign was for data to be taken once per
orbit for a period of approximately 1500s. About half way through the
campaign this strategy changed to observations once every other orbit,
but with a longer exposure time. Due to this strategy, different
binnings, in addition to providing differing time resolutions, result
in light curves with a differing proportion of gaps.  We have
therefore used a number of different binnings for the light curves in
our analysis, depending on the purpose for which it was employed, but
generally we either use the minimum time resolution for \xte\ PCA
Standard-2 data (16s), or the approximate orbital time scale of \xte\
(5760s). The exception is in the cross-correlation analysis, for which
we use an intermediate time bin width (256s), in order to explore
short time scale lags.

\begin{figure}
\epsscale{0.8}
\plotone{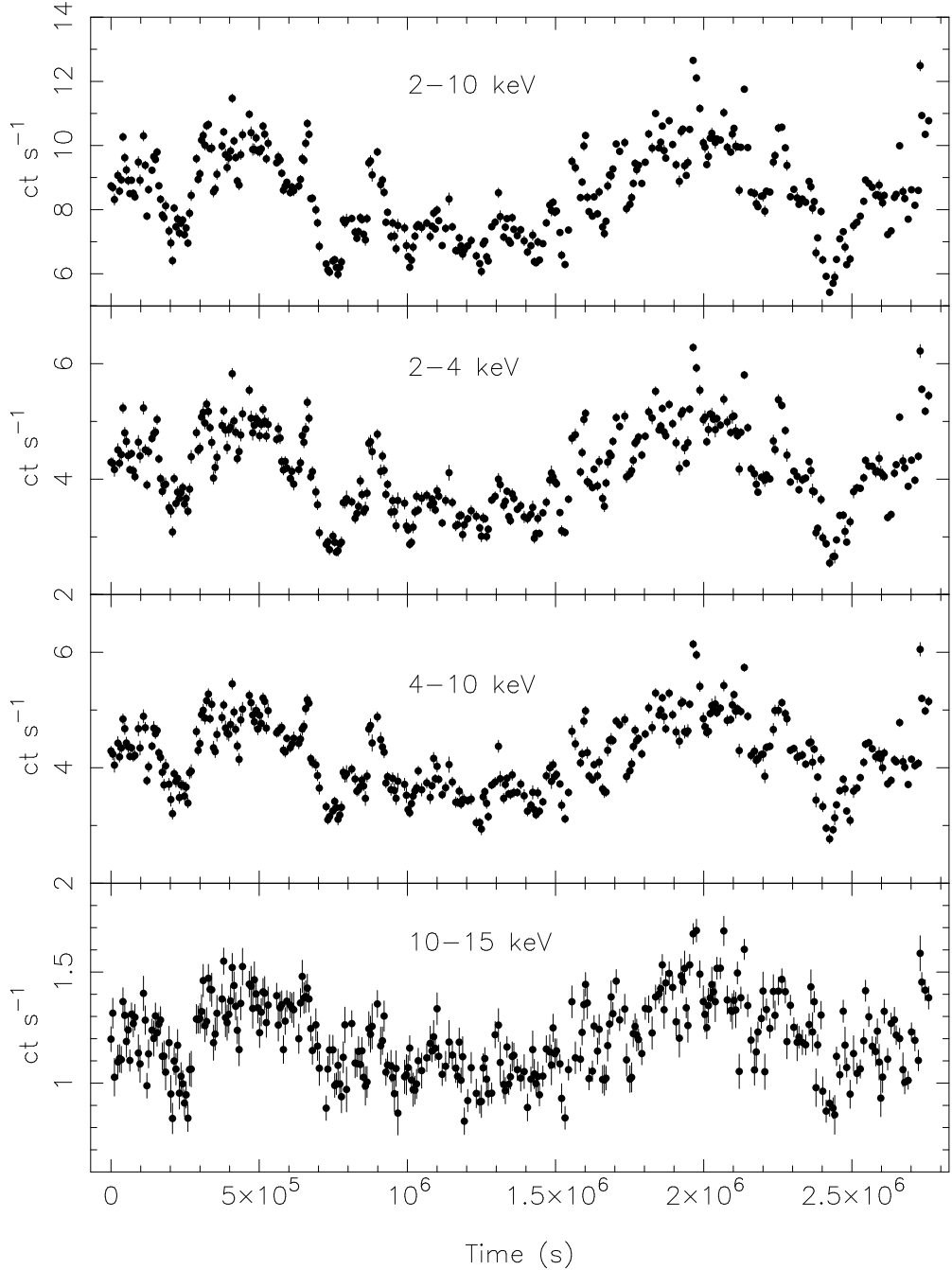}
\caption{X-ray light curves from the \xte\ PCA in (top to bottom) the
2-10 keV, 2-4 keV, 4-10 keV and 10-15 keV bands. Bins are 5760s -
approximately one orbit. For tidiness, only data bins which were at
least 10 per cent full are plotted here, although all data were used
in the analysis. The 2-10 keV light curve shows RMS variability of 15.9\%.
The 2-4, 4-10 and 10-15 keV data show amplitudes 
17.3\%, 14.9\% and 13.4\%, with the decrease in amplitude as a function
of energy being statistically significant (see Table~\ref{tab:var}).
The 15-24 keV light curve (not shown) does not follow this trend, with
RMS of 24.9\%, but this light curve seems to be dominated by residual
background variations, whose effect we explore in an Appendix.
\label{fig:lc}}
\end{figure}

\section{LIGHT CURVES}

Figure~\ref{fig:lc} shows the background-subtracted light curves in
four energy bands: 2-10 keV, 2-4 keV, 4-10 keV and 10-15 keV.  We will
occasionally refer to the last three ranges as the soft, medium and
hard bands respectively. All bands show significant variability, but
it is also apparent that there are some differences between the
bands. As a first attempt to quantify these differences, we have
computed the ``excess variance'' $\sigma^{2}_{\rm RMS}$ (e.g. Nandra
et al. 1997) from light curves with 16s and 5760s time bins for the
whole light curve. The mean and standard deviations of the light
curves, and their excess variances on these time scales, are shown in
Table~\ref{tab:var}. Note that with a ``red noise'' PSD, as exhibited
by NGC 7469 (see below), these RMS values should be dominated by the
longest-time scale variations (i.e. 10s of days). The amplitude of
variability decreases with increasing energy band, with the soft band
showing an excess variance corresponding to $17.7$~per cent RMS
variability, but the 10-15 keV band only showing $13.5$~per cent
variability.  If we calculate the excess variance using 16s time bins,
the corresponding RMS values are larger, $18.4$~per cent and
$15.6$~per cent, but maintain the trend. The larger percentage
increase in the harder energy band indicates that there may be more
power in short time scale flickering in the hard band. We return to
this in our analysis of the power spectra below.

\begin{figure}
\epsscale{0.8}
\plotone{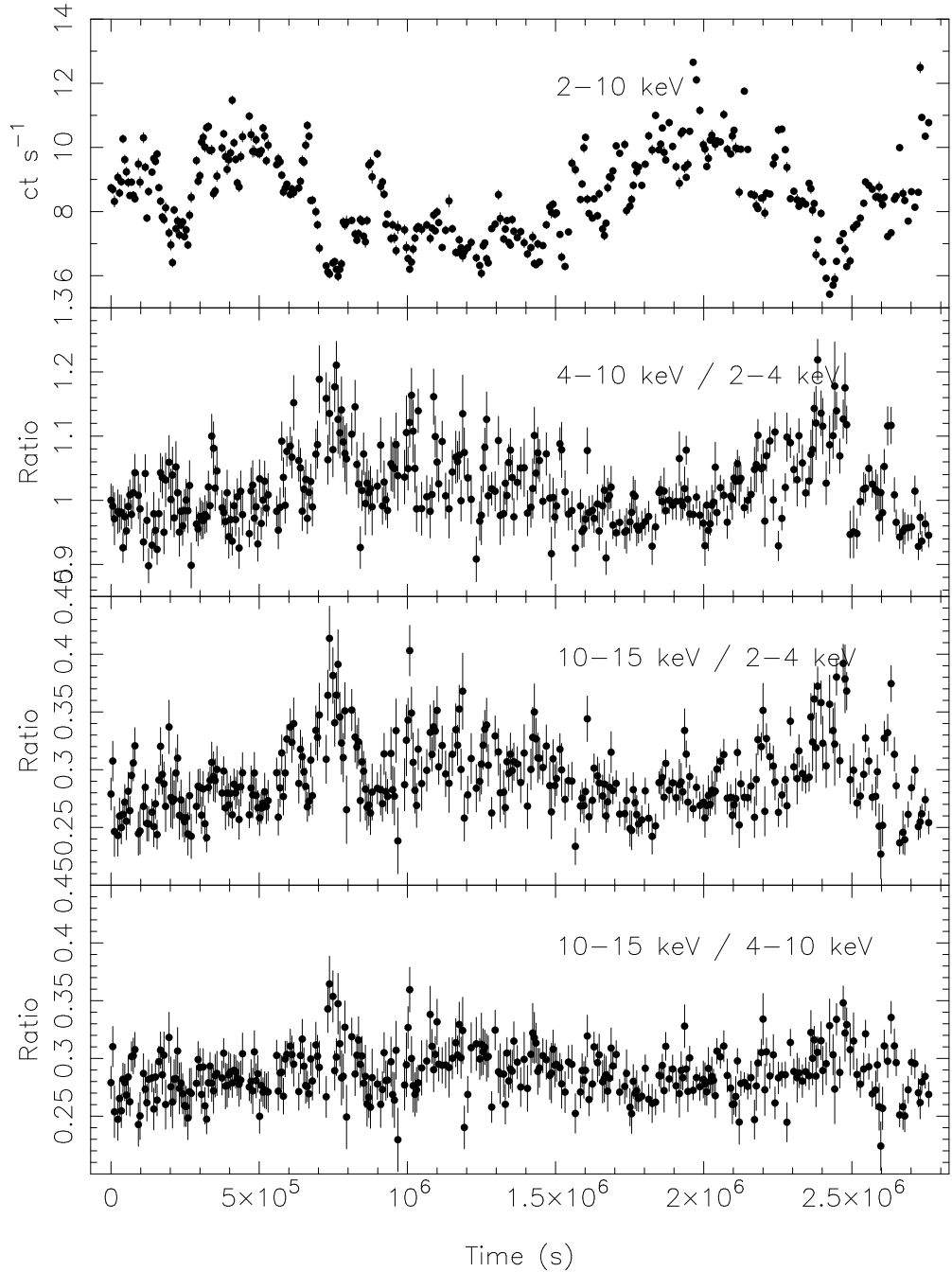}
\caption{X-ray light curve in the 2-10 keV band together with the
hardness ratios of three bands with an approximately-orbital bin size
of 5760s. All hardness ratios show significant variability with
reduced \chisq= 3.43, 3.46 and 1.61 (top to bottom).  The changes are
not strongly related to the 2-10 keV flux but there are sharp
increases in the hardness during the ``dips'' in flux around $7 \times
10^{5}$s and $2.4 \times 10^{6}$s. Changes in hardness can be very
rapid, with highly significant changes sometimes between orbits. 
\label{fig:hr}}
\end{figure}

The amplitude of the variations is apparently not the only difference
between the bands. There are strong changes in hardness ratio
(Fig.~\ref{fig:hr}) that seem to be related somewhat to the brightness
of the source, but not in a simple manner. Though fairly subtle, the
spectral variations can be very rapid, and can be characterized by a
distinct hardening of the spectrum during rapid dips in the flux.  The
hardness ratio changes can be extremely rapid - indeed significant
spectral variability is observed between adjacent 5760s bins.

\begin{figure}
\epsscale{1.0}
\plotone{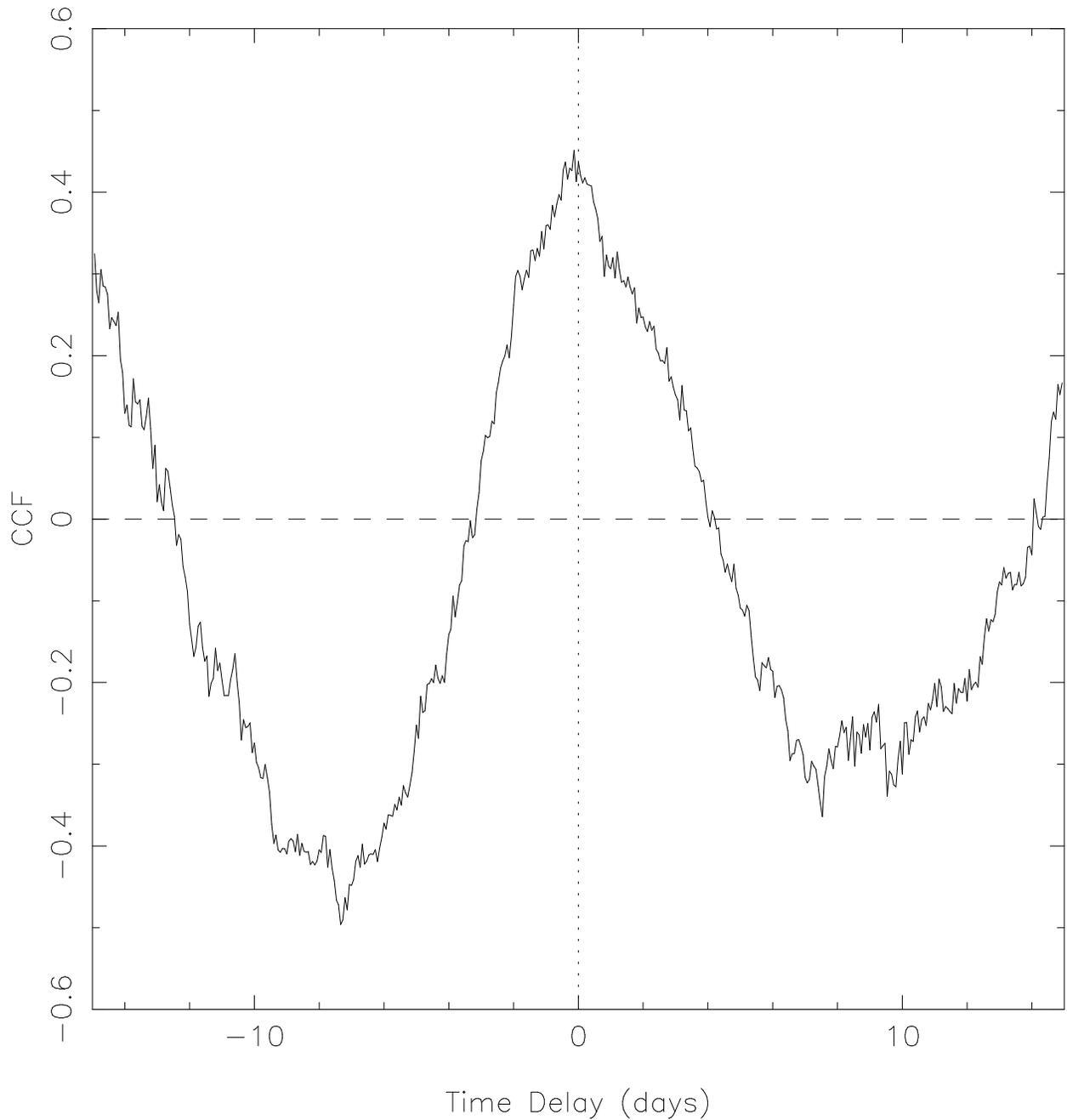}
\caption{Cross-correlation between the UV flux and the ratio of the
2-4 and 4-10 keV fluxes, calculated using the interpolation (ICCF)
method. The CCF has a peak at zero lag and Monte Carlo analysis
(Peterson et al. 1998) limits the lag to within $\sim 8.5$hrs of
zero. The peak correlation improves, however, if the hardness ratio is
calculated using longer time bins, indicating rapid changes in the
hardness ratio which are not seen in the UV flux.
\label{fig:uvhr_ccf}}
\end{figure}

Spectral variability has already been reported for this source by
Leighly et al. (1996) and was studied in detail by N2K.  They found
that the X-ray spectral index was variable, and correlated with the
ultraviolet flux, but not the X-ray flux. Fig.~\ref{fig:hr} confirms
this, in that the hardness ratios show no obvious correlation with the
X-ray flux. We do find a correlation between the hardness ratio and
the ultraviolet flux, however. Indeed, as we can determine the
hardness ratio with a lower exposure time than the spectral index,
this may be useful in determining the time lag between the bands.
Fig.~\ref{fig:uvhr_ccf} shows the cross correlation between the
ultraviolet 1315\AA\ light curve (from Kriss et al. 2000; see also
Wanders et al. 1997) with full sampling, and the 2-4 keV/4-10 keV {\it
softness} ratio in orbital (5760s) bins. This was calculated using the
interpolation or ICCF method (Gaskell \& Peterson 1986; White \&
Peterson 1994).  We choose the softness ratio as this should correlate
positively with the UV at zero lag. There is a peak in the correlation
function the light curves close to zero lag.  Using the method
of Peterson et al. (1998), the best limit we obtain on the time lag
between the UV and the softness ratio is $-2.2 \pm 8.4$ hours,
consistent with zero lag. We note, however, that the peak correlation
coefficient is much lower than that obtained by N2K for the UV
vs. Gamma (r=0.81). Furthermore, if we bin the hardness ratio in 0.5d
bins, instead of orbital bins, we obtain a much higher zero lag
correlation of r=0.76. Although some of this is undoubtedly due to
larger errors in the hardness ratio for smaller time bins this is very
unlikely to account for the whole effect. Some of the reduction in the
correlation with the smaller bins appears to be due to the fact that
the hardness ratio shows very rapid changes which are not
observed in the UV light curve. In particular we see very clear
changes in hardness ratio from orbit to orbit (i.e. in 90 minute
intervals), and the UV light curve never shows variations this fast
(Wanders et al. 1997).

\begin{figure}
\epsscale{0.8}
\plotone{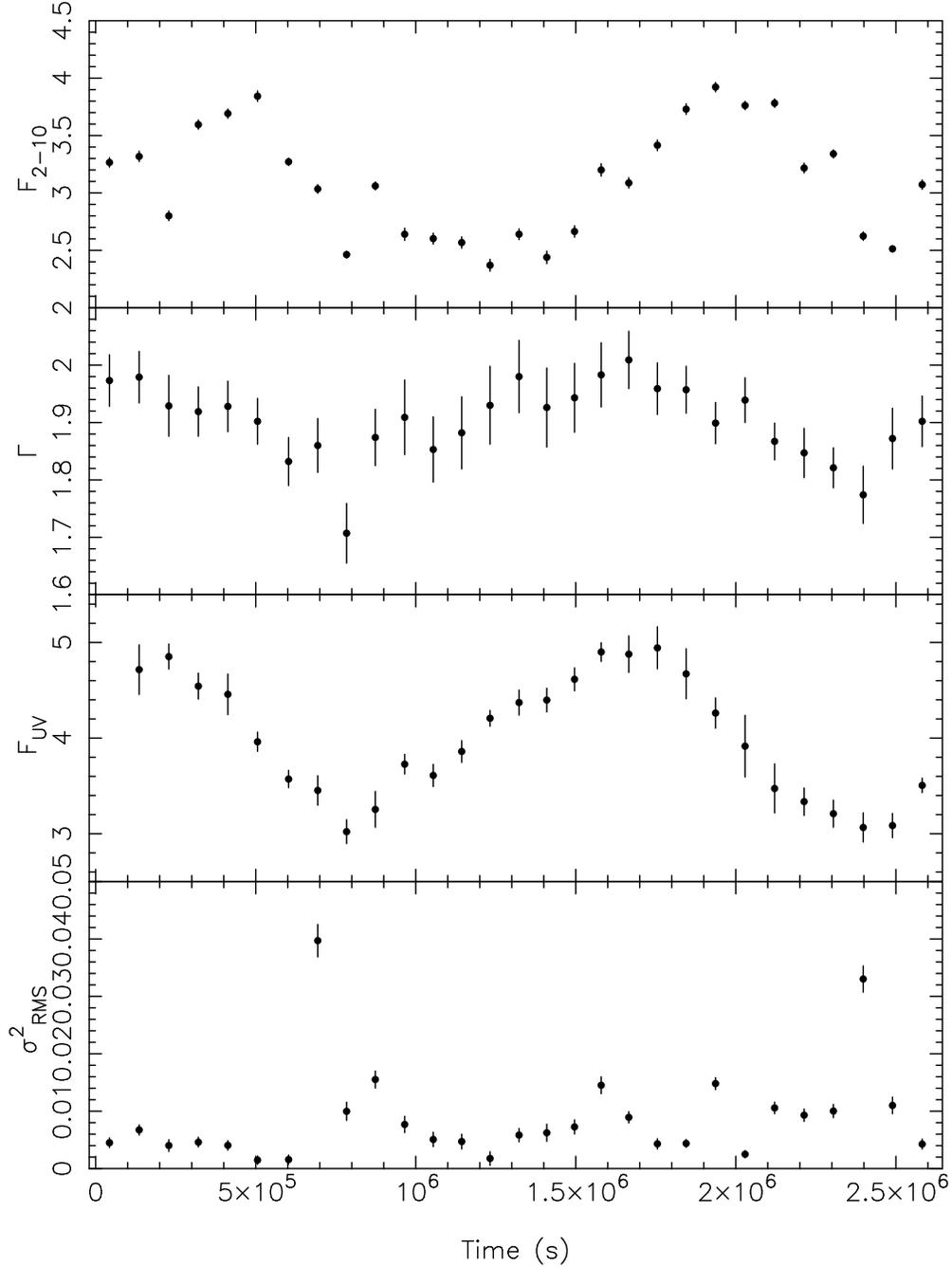}
\caption{Light curves in approximately daily bins of (descending) the
2-10 keV X-ray flux, power law photon index (2-20 keV) and ultraviolet
flux at 1315\AA\ (all from N2K) together with the excess variance of
the 2-10 keV light curve calculated in daily bins. Very significant
changes are seen in the excess variance. These are not obviously
correlated with the flux, although there are two sharp increases in
the variance around the times when the X-ray spectrum is hardest, and
the UV flux the weakest. Note also the smoothness of the UV light
curve on short time scales, compared to that of the hardness ratio
(Fig.~\ref{fig:hr}).
\label{fig:rms_lc}}
\end{figure}

Our dataset is sufficiently large that we have been able to examine the
light curves for changes in excess variance over the period of the
observation. We examined the 16s light curves, and calculated the
excess variance in these for each day of the observation separately.
The results for the 2-10 keV light curve are shown in the bottom panel
of Fig.~\ref{fig:rms_lc}, together with the the light curves of the
X-ray flux, X-ray photon index and UV flux (N2K). Periods of enhanced
X-ray variability seem to coincide with, or just precede, the times
where the X-ray spectrum is hardest and the UV flux weakest
(interpreted by N2K as being due to a corona with high
temperature). It is difficult to assess the statistical significance
of the relationship between the change in variability parameters and
the spectral state but it is nonetheless an intriguing possibility which
we discuss further below.

\begin{figure}
\epsscale{0.8}
\plotone{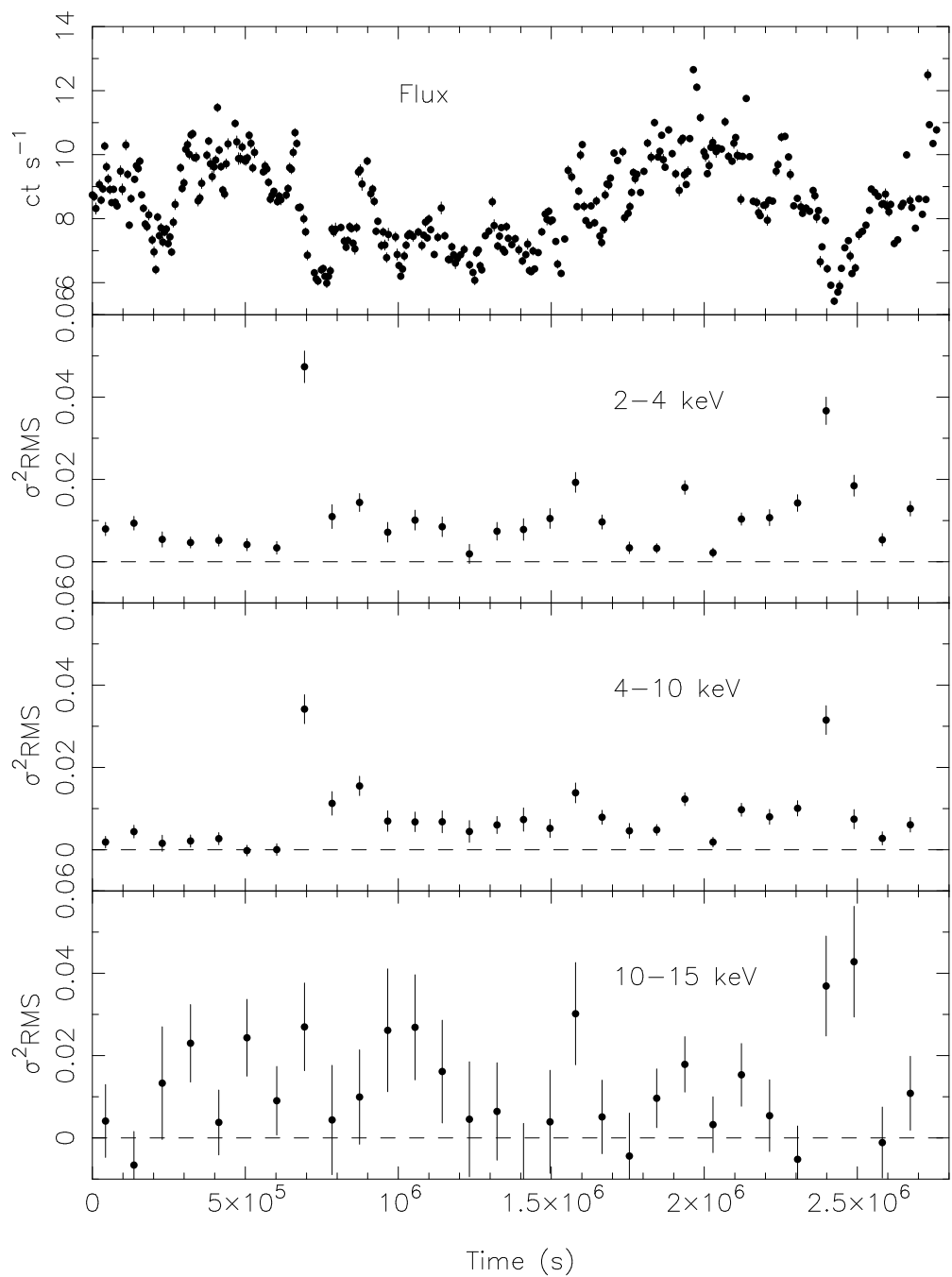}
\caption{2-10 keV X-ray light curve in orbital bins (top panel),
together with the excess variance in daily bins in (descending) the
2-4, 4-10 and 10-15 keV bands. The excess variance in all bands shows
changes, and in particular in the 2-4 and 4-10 keV bands seems to show
peaks at the points where the X-ray flux reaches a minimum, similar to
the 2-10 keV variance (Fig.~\ref{fig:rms_lc}). There is no
simple correlation with the flux however.
\label{fig:rms}}
\end{figure}

We also calculated the excess variance in daily bins for the narrow
bands, using the 16s light curves. The results are shown in
Fig.~\ref{fig:rms}. The mean excess variances found on 1-day time
scales were all similar, with the 2-4, 4-10 and 10-15 keV bands
exhibiting $8.8^{+0.2}_{-0.2}$ per cent, $7.5^{+0.2}_{-0.3}$ per cent
and $9.5^{+1.1}_{-0.8}$ per cent respectively.  We contrast this with
the variance over the entire 30 day period, which shows a clear
decrease with energy band. This once again indicates relatively more
power on the short time scales in the hard band. In no band, however,
is the excess variance consistent with a constant over the period of
the observation. As expected, the 2-4 and 4-10 keV variance show
similar patterns to the 2-10 keV (Fig.~\ref{fig:rms_lc}). The behavior
in the 10-15 keV band is somewhat less clear in this regard, however.

\section{CORRELATION FUNCTIONS}

\begin{figure}
\epsscale{0.8}
\plotone{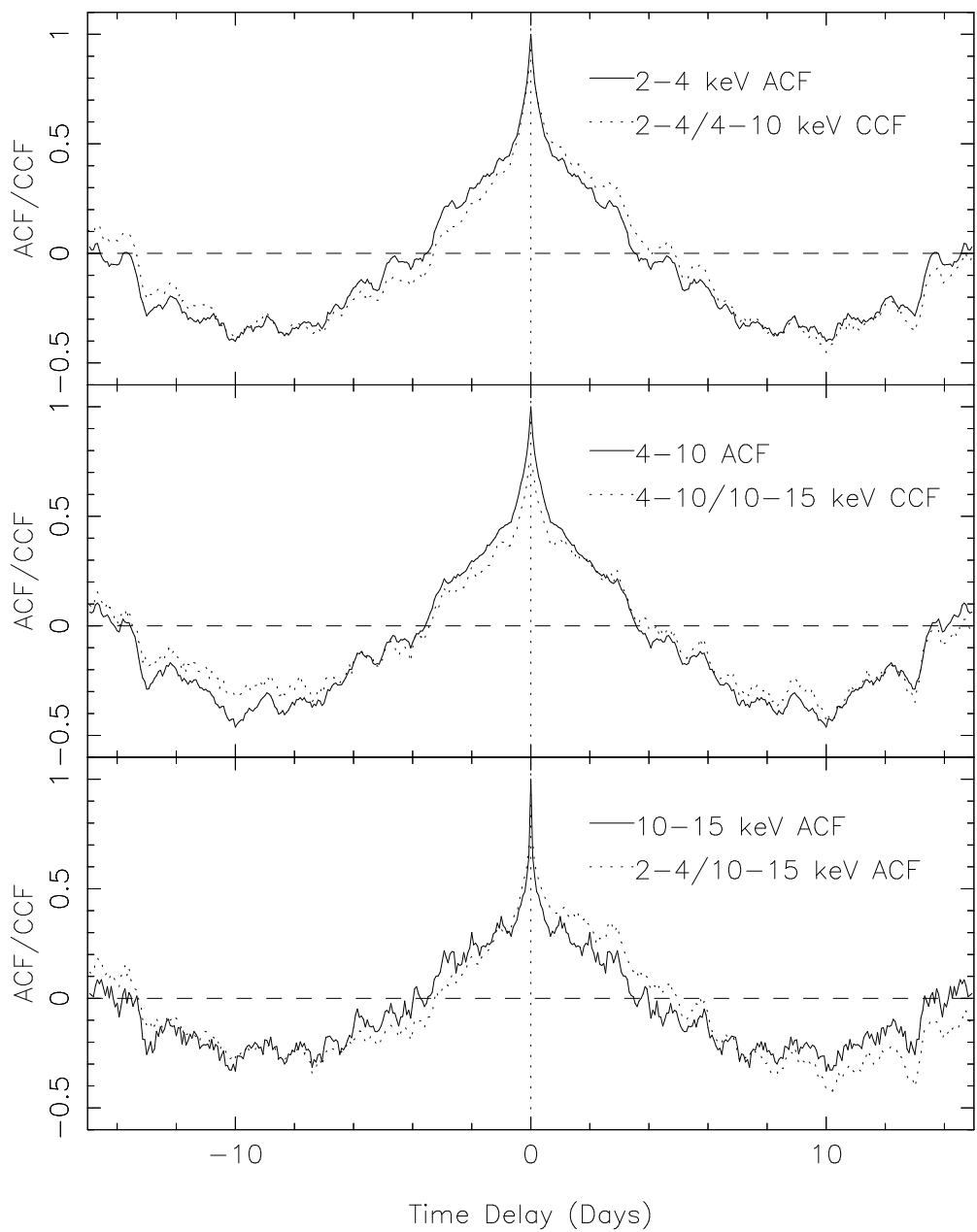}
\caption{Auto-correlation functions of the X-ray light curves in the
2-4, 4-10 and 10-15 keV bands (solid lines, descending). Zero points
have been suppressed. The width at zero intensity is very similar in
all bands. Also shown on this plot are various cross-correlation
functions (see also Fig.~\ref{fig:ccf}) plotted as dotted lines. The
CCFs are clearly skewed to positive lags compared to the (symmetric)
ACFs.  This plot also demonstrates that at positive lags of a few
days, the 2-4 keV light curve is better correlated with both the 4-10
and 10-15 keV curves than it is with itself.
\label{fig:acf}}
\end{figure}

Fig.~\ref{fig:acf} shows the autocorrelation function in the three
narrow energy bands, using the ICCF method. The ACFs are all very similar,
with a strong and relatively narrow peak but broader ``wings'' on a
time scale of a few days. They all reach zero at a time delay of
around 4d. Within a delay of $\pm 5$~d the ACFs can be fit with a
double gaussian model, with widths ($\sigma$) for the 2-4, 4-10 and
10-15 keV ACFs of 0.21, 0.23 and 0.17 days and 1.72, 1.70 and 1.78
days for the narrow and broad parts of the ACF. 

\begin{figure}
\epsscale{0.8}
\plotone{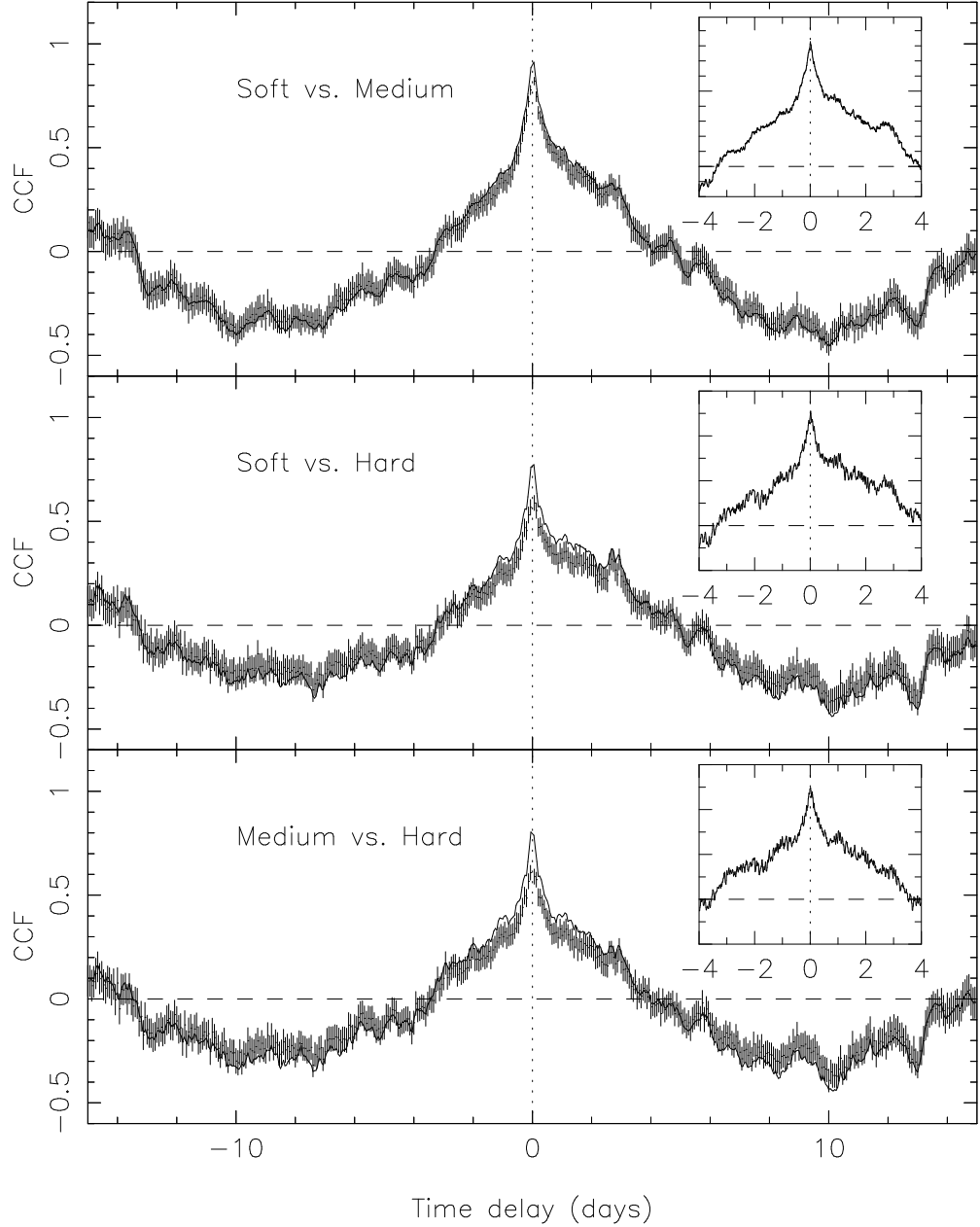}
\caption{Cross-correlation functions for the X-ray light curves. The
lines represent the interpolation cross-correlation (ICCF) and the
points with error bars the Z-transformed Discrete Correlation
Function.  There is reasonable agreement between the two methods. In
all cases the softer light curve is taken as the ``driving'' light
curve, meaning that positive lags correspond to the soft X-ray
variations leading the hard X-rays. The main panels are (descending)
the 2-4/4-10 keV, 2-4/10-15 keV and 4-10/10-15 keV CCFs calculated
with a bin size of 5760s. The inset panels show the CCFs calculated
with a bin size of 256s and limited to lags of $\pm 5$d. All CCFs show
a narrow peak (FWHM$\sim 0.5d$) consistent with zero lag. This is
superimposed on broader ``wings'' (FWHM$\sim 5$d) which are shifted to
positive lags of $\sim 0.5$d. Peak correlation coefficients and limits
on time lags are shown in Table~\ref{tab:ccf}.
\label{fig:ccf}}
\end{figure}

We have also calculated the cross-correlation functions between the
bands, which are shown in Fig.~\ref{fig:ccf}. Here we have also
employed the Z-transformed discrete correlation function or ZDCF
(Edelson \& Krolik 1986; Alexander 1997). While the general forms of
the ZDCF and ICCF are similar, at small lag there is some disagreement
in the sense that the ICCF shows a stronger correlation. The reason
for this is unclear, but henceforth we adopt this and quote results
only for the ICCF method. This has the advantage of allowing us to
employ the method of Peterson et al. (1998) to examine uncertainties
and limits on any time lags and in any case the difference is
small. To provide a broad view of the correlation between the bands on
$\sim$~day time scales, we first calculated the correlation functions
using the 5760s light curves, with the results being shown in the main
panels of Fig.~\ref{fig:ccf}. In all cases, we chose the lower-energy
light curve as the ``driving'' emission, such that a positive lag
would mean that the variations in the higher energy emission followed
after those in the softer band. The CCFs are in practice always peaked
very close to zero lag. Peak correlation coefficients and estimated
lags are shown in Table~\ref{tab:ccf}.  While the peak of the CCFs in
this analysis is clearly very close to zero, they are not
symmetric. The asymmetry is in the sense that the softer band leads
the harder one on time scales of a few days. This can also be seen
clearly in Fig.~\ref{fig:acf}, which shows that the asymmetry is
rather strong. Indeed we find that at positive lags of a few days, the
2-4 keV light curve is better correlated with the higher-energy light
curves than it is with itself. Interestingly, we can fit the $\pm 5$d
CCFs with the same model we applied to the ACFs. The double gaussian
fit results in very similar widths of 0.24, 0.21, 0.24 and 1.71 1.74
and 1.82 days for the narrow and broad components. Although the narrow
component peaks very close to zero lag, however, with centroids of
-0.006, -0.015 and -0.020 days for the 2-4/4-10, 4-10/10-15 and
2-4/10-15 keV CCFs, we find the broad component to be shifted to
positive lags by 0.51, 0.23 and 0.79 days. One interpretation of these
results is that there are two variability components or variability
processes, one of which produces predominantly short time scale
($<1$d) variations which are synchronous across the X-ray waveband,
and another which produces variability on $>1$d time scales which
shows time delays in the sense that the hard X-ray variations lag
those at soft X-rays by several tenths of a day. We return to this in
the discussion.

This orbital analysis is insensitive to lags shorter than the bin size
and we have therefore also performed cross-correlations using a
smaller bin size of 256s. These are shown in the inset to
Fig.~\ref{fig:ccf}. Peak lags and limits on the time delay are shown
in Table~\ref{tab:ccf}. The best limits are obtained for the 256s
light curves, which show that the variations are synchronous to within
a few tens of minutes - certainly less than 1hr. We note that, as
discussed by, e.g. Welsh (1999), there is some inaccuracy in both the
lag and error determinations due to the ``red noise'' nature of the
light curves. Our main conclusion that the CCFs are consistent with
zero lag is unlikely to be affected by this, however. The zero lag
correlations are also very strong, with negligible formal chance
probability. The analysis so far has demonstrated some similarities
and some difference in the variability behavior with time and spectral
properties. Arguably the most sensitive tool for the analysis of
variability is the power spectral density function, which we now go on
to explore.

\section{POWER SPECTRAL DENSITY}

\begin{figure}
\epsscale{0.8}
\plotone{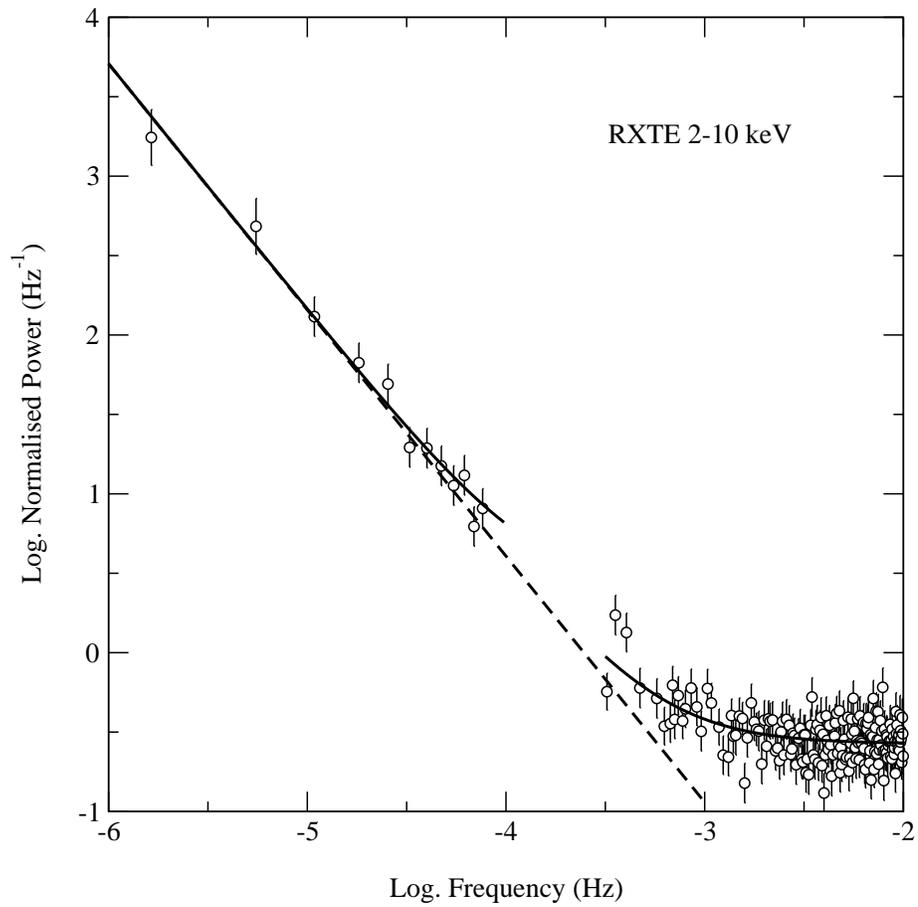}
\caption{The 2-10 keV PSD. Fitting of the low and high frequency
portions of the PSD together with a power law gives a good fit,
with a slope of $\alpha=1.55^{+0.10}_{-0.06}$ and a normalization
of $4.0 \pm 0.6$~Hz$^{-1}$. There are no clear features such as
spectral breaks or periodicities in the PSD, although there is 
some structure especially in the high frequency PSD, with a possible
excess of power at $\sim 10^{-3.5}$~Hz
\label{fig:pds}}
\end{figure}

\subsection{Method of calculation}

The RXTE light curve of NGC~7469 is unevenly sampled and so the
estimation of the power spectrum is not trivial. Here we estimate the
overall power spectral density function of the source in two
steps.  

Firstly, we used the 16-sec binned light curve to compute the
high frequency PSD. Although the light curve is not evenly sampled
there are 230 parts of it with length between 1000 and $\sim 4000$ sec
with no gaps in them.  Therefore, we can estimate the PSD for each
part and then average the resulting spectra to get a good estimation
of the PSD at high frequencies. For each part we computed the
periodogram using the equation,

\begin{equation}
\hat{I} (\nu_{j}) = \frac{ (\Delta t/N)
\{ \sum_{i=1}^{N}[x(t_{i})-\bar{x}]e^{-i2\pi \nu_{j} t_{i} }\}^{2}}
{ \bar{x}^{2} },
\end{equation}

where: $\Delta t=16$ sec, $\nu_{j}=j/(N\Delta t)$, $j=1,2,\ldots, N/2$,
and $\bar{x}$ is the mean count rate of each part. The periodogram, as
defined by equation (1), is normalized to the mean count rate square and
has units of $Hz^{-1}$. With this normalization, integration over
positive frequencies yields half of the light curve's relative variance.
This normalization is necessary for the comparison between PSD in
different energy bands.

Since the 230 parts of the light curve do not have the same length,
the frequencies at which their periodograms are computed are not
exactly the same. To combine them, we sorted them in order of
increasing frequency, calculated their logarithm, grouped them into
bins of $20$ and computed their average value at each bin. These
are our final estimates of the logarithmic PSD at the geometric mean
frequency of each bin (Papadakis \& Lawrence 1993a).

To estimate the low frequency PSD we used the 5760s binned light
curves. The use of the approximate RXTE orbital period as the bin size
results in an evenly sampled light curve with very few missing points
(74 out of 480, $\sim 15$\% of the total). Note that these points are
randomly distributed over the whole light curve, with no more than two
points in a row missing. We accounted for them using linear
interpolation between the two bins adjacent to the gaps, adding
appropriate Poisson noise. We used equation (1) to compute the
periodogram for the 5760 sec binned light curve. As above with the
high frequency PSD, we computed the logarithm of the periodogram and
averaged it in groups of size 20 (Papadakis \& Lawrence 1993a).

Aside from the interpolation, another potential problem with the
binning process is that, because the original 16s light curve also has
gaps, the binning process results in data points with different
exposure times.  Therefore, the mean count rate at each bin is the
result of averaging the signal over segments of different length at
each bin. Furthermore, the location of the segment within each bin
will not be the same for all bins - therefore, although we accept the
average count rate at each bin as representative of the source count
rate at the center of the bin, in some cases it will actually
correspond to a time point other than the bin center. Consequently,
strictly speaking, the 5760 sec binned light curve is not a random
realization of the parent, source process and the binning may alter
the light curve's properties. We believe, however, that neither the
interpolation nor the binning process will have a significant effect
on our results. The source PSD is rather steep with $\alpha>1.5$ as
shown by the high frequency PSD alone.  Both the interpolation and the
binning effect the highest frequencies most, where there is less
power, and we therefore do not believe these effects will bias our
results severely. Note that the best fitting slope of the PSD is
mainly determined by the low frequency points. Furthermore, the
effects of binning and interpolation should have the same affect on
the PSD regardless of the energy range considered, and therefore our
comparison of PSDs in different energy bands discussed below will not
be affected.

\subsection{2-10 keV PSD}

%\begin{figure} 
%\plotone{./figs/fig_pds_lc.ps} 
%\caption{Light curve of the
%PSD parameters in 3 day time bins. The top panels show the parameters for
%a power law fit with both the index and normalization free. The bottom
%panel shows the normalization assuming the shape (slope) is fixed at the
%mean value (indicated by the dashed lines in the first panel). Although it
%is not possible to determine whether the shape, normalization or both
%changes with time, clearly at least one of them does, and the bottom panel
%shows behaviour similar to that of the RMS, in that the normalization
%increases significantly around days 8 and 28, where the RMS shows dramatic
%increases (Fig.~\ref{fig:rms_lc}).  
%\label{fig:pds_lc}}
%\end{figure}

Fig.~\ref{fig:pds} shows the PSD in the 2-10 keV band calculated by
this method. Several things are apparent from this. First, there are
no obvious ``features'' in the PSD, such as a low frequency knee or
high frequency break. Neither are there any clear periodicities
evident.  We have used standard $\chi^{2}$ statistics to fit different
models to the spectrum, evaluating the 68 per cent confidence regions
for the model parameters using the prescription of Lampton, Margon \&
Bowyer (1976). The first model was a power law plus a constant of the
form:

\begin{equation}
P(\nu)=A(10^{4}\nu)^{-\alpha} + C,
\end{equation}

where: $A$ is the amplitude (in Hz$^{-1}$) at $10^{4}$ Hz, $\alpha$ is
the slope and $C$ is a constant, which represents the constant Poisson
power level. Rather than letting $C$ be free during the model fitting,
we fixed its value to the expected noise power level. This matches the
apparent noise power of the data very well indeed, as can be seen at
high frequencies. This model was fitted to both the low and high
frequency PSD of the source, allowing the slope and the amplitude to
be different for the two parts of PSD. The results are listed in
Table~\ref{tab:pds_sep}. A power law is a good fit to both spectra and
the two spectra give consistent indices and amplitudes within the
errors.

We also fitted this model to the combined low and high frequency power
spectra, with the results shown in Table~\ref{tab:pds_joint}.  The
power law index of $\alpha=1.55^{+0.10}_{-0.06}$
(Table~\ref{tab:pds_joint}) is very similar to the typical value
inferred by Lawrence \& Papadakis (1993) from the analysis of \ex\
data of a few Seyfert 1 galaxies. While the fit is formally
acceptable, the PSD does seem to show some structure, especially at
high frequencies. One interpretation of this is that there is a
``QPO''--like feature at $\sim 10^{-3.5}$ Hz. If we add a Gaussian to
the power law model we find a $\chi^{2}$ of 546/571, a reduction of
45.5 for 3 dof. This is formally of very high significance based on
the F-test ($>99.9$~per cent confidence). The slope and amplitude do
not change much ($\alpha=1.59$ and $A=4.1$
cf. Table~\ref{tab:pds_joint}), and the centroid of the ``QPO'' is:
$\sim 3.8\times10^{-4}$ Hz, the width (sigma) is: $\sim 2.2\times
10^{-5}$ Hz, and the amplitude $\sim 0.7$ Hz$^{-1}$.  Despite the high
significance, we are cautious about claiming the detection of a
``QPO'' in NGC 7469 for several reasons. First, the feature is very
close to the frequency gap between the low and high frequency PSD,
meaning we cannot determine the shape of the feature, and therefore be
sure that it is in fact indicative of a periodicity. The ``QPO'' in
the 2-10 keV PSD relies on the accuracy of the lowest point in the
high frequency PSD. It is therefore difficult to know whether the poor
fit -- or the improvement when the gaussian is included -- is due to
some kind of continuum feature or break. We note that the PSD of Cyg
X-1 shows high frequency structure also (Nowak et al. 1999) and we may
simply be misinterpreting this as some kind of quasi-periodic signal.

The changes in RMS variability parameter shown in
Fig.~\ref{fig:rms_lc} imply that there is some change in the PSD with
time, as the RMS represents an integral of the PSD over the frequency
ranges sampled by the relevant light curve portion. For the daily
calculations (Fig.~\ref{fig:rms_lc} and Fig.~\ref{fig:rms}) the RMS
should represent approximately the integral of the PSD above a
frequency of $\sim 10^{-5}$~Hz. This could represent either a change
in the shape of the PSD, a change in its normalization or some
combination of the two. We have investigated this by calculating the
PSD as a function of time. To sample sufficient variability to provide
decent constraints, we were only able to calculate the PSD every 3
days, and characterized this as a power law.
%The results are shown in Fig.~\ref{fig:pds_lc}. 
Unfortunately this does not able us to distinguish unambiguously the
origin of the changes in the PDS shape during the UV ``dips''. When
the shape (slope) is held constant at the mean value, however, the
normalization shows clear changes, consistent with the changes in RMS
variability parameter seen in Fig.~\ref{fig:rms_lc}. We cannot rule
out the alternative of a change of shape, however. 

\subsection{PSD as a function of energy}

\begin{figure}
\epsscale{0.8}
\plotone{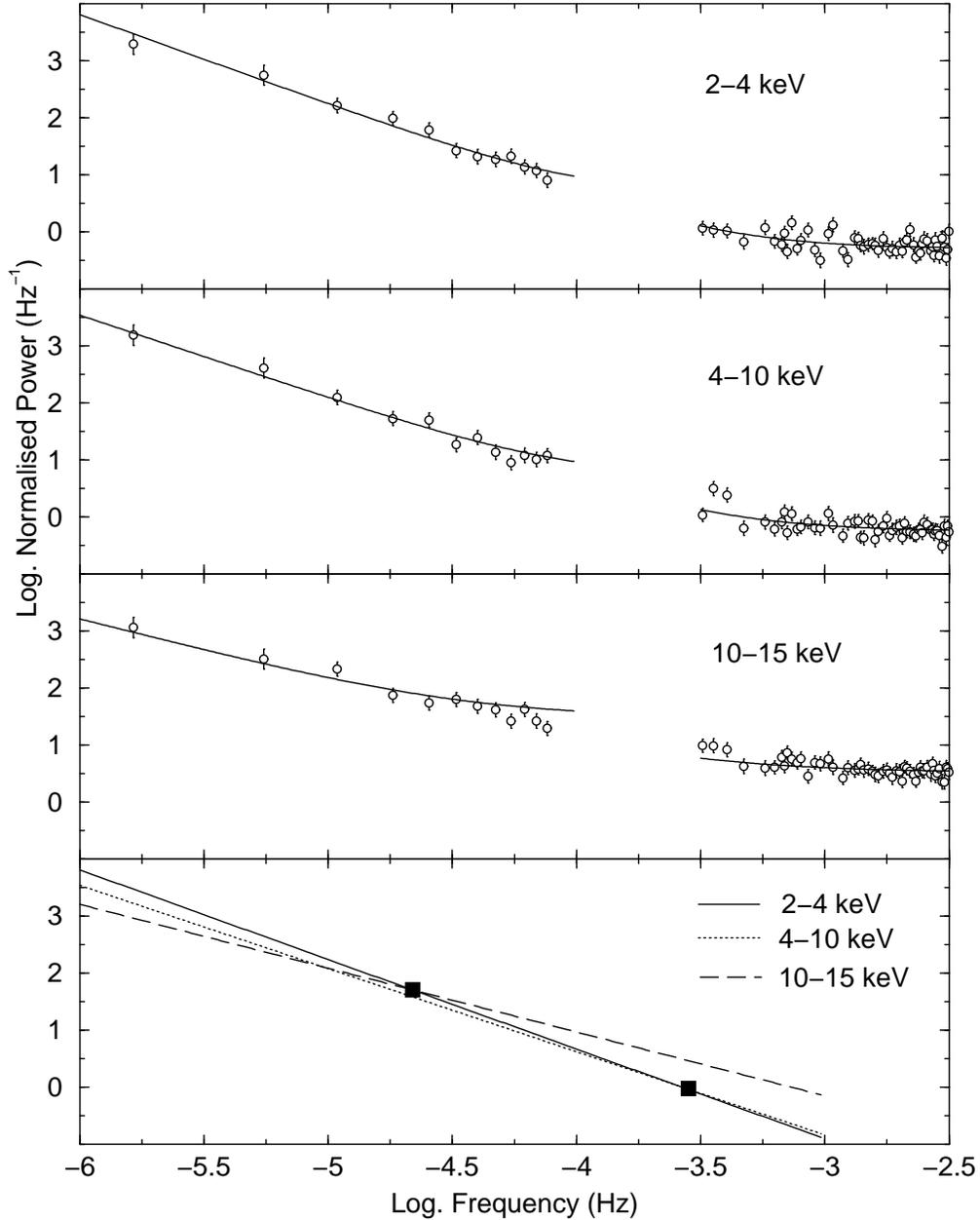}
\caption{The top three panels show the power spectral density in the 3
different energy bands. They have been fitted with a power law model
with a fixed constant representing the Poisson noise (solid
lines). Note that the low and high frequency PSDs have different
Poisson noise levels, however.  The dashed lines show the power law
only. The PSD shows a ``hardening'' as a function of energy, with the
10-15 keV PSD being significantly flatter than that at lower energies
(see also Table~\ref{tab:pds_joint}).  There is also some
``structure'' in the high frequency PSD, especially visible in the
higher energy bands which can be fitted, albeit not uniquely with a
``QPO''-like feature. The best-fit power law models for the power density 
spectra are shown in the bottom panel. This shows that
The low energy PSD shows most power on the longest time scales, but
that the opposite is true at high frequencies. The crossover point
seems to be around $10^{-5}$~Hz or around 1d.  
\label{fig:3band_pds}}
\end{figure}

%\begin{figure}
%\epsscale{0.8}
%\plotone{./figs/fig_pds_mdl.ps}
%\caption{
%Best-fit power law models for the power density spectra
%in the three bands shown in Fig.~\ref{fig:3band_pds}. This shows that
%the low energy PSD shows most power on the longest time scales, but
%that the opposite is true at high frequencies. The crossover point
%seems to be around $10^{-5}$~Hz or around 1d.  
%\label{fig:pds_mdl}}
%\end{figure}

We have calculated the PSD in the 3 narrow energy bands
(Fig.~\ref{fig:lc}) of 2-4, 4-10, 10-15 keV by the same method as the
2-10 keV PSD. The results are shown in Fig.~\ref{fig:3band_pds}. The
results of fitting the high and low frequency portions separately are
also shown in Table~\ref{tab:pds_sep}. A good fit is obtained to all
the spectra. The low and high frequency slopes are consistent within
the errors for each individual energy band. In the 4-10 and 10-15 keV cases,
however, the spectra show a systematically higher normalization in the
high frequency spectra. Such an effect could be observed if there is a
high frequency break in the PSD (e.g. Nowak \& Chiang 2000). This
break would have to occur above $10^{-4}$~Hz but it is not clearly
observed in the high frequency PSD. This leaves the rather unlikely
possibility that the break occurs precisely in the gap between the two
PSDs. It seems equally unlikely that the PSD shows a ``step'' in power
in this gap, maintaining (approximately) the same slope. It seems more
likely that this apparent discrepancy in normalization is due to the
structure or feature previously noted in the 2-10 keV PSD. Indeed
visual inspection of the PSDs show that the strength of this
``feature'' increases with energy band.

Whatever the nature of the structure in the high frequency power
spectrum it is clear from our analysis that the PSD flattens (or
``hardens'' to use a term borrowed from the energy spectra) as a
function of energy. This is most clearly demonstrated when the low and
high frequency PSDs are fitted jointly, with the results for a power
law model shown in Table~\ref{tab:pds_joint}. The slope steepens from
$\alpha=1.57^{+0.10}_{-0.08}$ in the 2-4 keV band to
$\alpha=1.12^{+0.10}_{-0.08}$ in the 10-15 keV band. The best fit 4-10
keV PSD slope lies between the two but is formally consistent with the
2-4 keV, though inconsistent with the 10-15 keV slope. The normalizations
of all three PSDs (at $10^{-4}$~Hz) are all consistent within the
errors. 

Despite the apparent structure at high frequencies, the fits are
formally acceptable at $< 90$~per cent confidence, with no clear
evidence for high or low energy breaks. Although the power law is
probably only a crude parameterization of a more complex form, this
very clearly shows there are differences in the PSDs, and the
difference is in the sense that the high frequency power spectrum
shows more power at high energies. As the PSDs are normalized
correctly, Fig.~\ref{fig:3band_pds}, also shows that the opposite is
true at low frequencies, i.e. that there is more power on long time
scales at softer energies.  This is demonstrated best by the bottom
panel of Fig.~\ref{fig:3band_pds}, which shows the best-fit power law
models for the 3 PSDs.  There appears to be a ``pivot'' point just
above $10^{-5}$~Hz where the normalized power is roughly equal. The
soft band has more power below this frequency, and the hard band more
above. This behavior is similar to that seen in Cyg X-1 (Nowak et
al. 1999) and has strong implications for the emission mechanisms.

\section{DISCUSSION}

We have presented a detailed temporal analysis of the 30d RXTE
observation of NGC 7469, which was simultaneous with IUE. Strong
variability was observed in all X-ray bands, with the variability
amplitude observed over the whole observation being a decreasing
function of energy. The variability amplitude calculated on daily time
scales is similar in all energy bands, however, and shows significant
changes, indicating non-stationarity. The changes are such that the
most pronounced variability occurs when the X-ray spectrum is hardest
and the UV flux weakest. The UV flux is positively correlated with the
spectral softness ratio, and the cross-correlation shows a strong peak
at zero lag, constraining the lag to be less than $\sim
8.5$~hours. The softness ratio changes are more rapid than those in
the UV, however. The autocorrelation functions are very similar in the
2-4, 4-10 and 10-15 keV bands, all showing a sharp peak with FWHM of
about 0.5d and broader ``wings'' with FWHM $\sim 5$d, reaching zero at
a time scale of $\sim 4$d. The cross-correlations show a strong peak
at zero lag, but are asymmetric, towards the side where the hard
X-rays lag. They can be interpreted as having the same two components
as the ACFs, but with the 5d FWHM component being delayed by several
tenths of a day, in the sense that the soft X-ray variations lead
those in the hard.  The 2-10 power density spectrum shows no low
frequency turnover or high frequency break. There is marginal evidence
for a feature at a frequency of $3.8 \times 10^{-4}$~Hz, but this may
simply be due to structure or complexity in the high frequency
PSD. The most striking result of the PSD analysis is that the 10-15
keV PSD is clearly flatter than the 2-4 and 4-10 keV PSD.  In other
words the PSD flattens or ``hardens'' as a function of energy. We now
discuss these results in terms of the models for NGC 7469 and AGN in
general.

\subsection{Thermal Comptonization models}

Considerable evidence is emerging that the dominant radiation
mechanism producing X-rays in AGN is thermal Comptonization.  The
copious optical/UV radiation field is thought to act as the seed
photons, which are upscattered by a corona of very hot plasma with
temperatures of several hundred keV. As mentioned in the introduction,
such observations fit the X-ray spectra of AGN well. Variability data
also support such an interpretation. In the simplest such models, we
expect to see correlated variations in the optical/UV seed and the
X-rays, perhaps with a time lag in the sense that variations in the
lower energy emission precede those at higher energies.  Correlations
between the UV and X-ray have have been observed in some cases (Clavel
et al. 1992; Edelson et al. 1996). Strong correlations have also been
observed between the UV and EUV emission (Marshall et al. 1997) and
between the EUV and the X-ray emission (Uttley et al. 2000), and a
time lag has been claimed in one case (Chiang et al. 2000). Other
observations have shown much less clear behavior both in terms of the
correlations and the lag (Done et al. 1990; Maoz et al. 2000; Edelson
et al 2000), including those of NGC 7469 (N98).

In a realistic corona a simple correlation plus lag is not necessarily
expected, however. As the UV seed flux cools the corona, this can
change the coronal properties and affect the X-ray spectral shape
(e.g. Pietrini \& Krolik 1995; Haardt, Maraschi \& Ghisselini 1997;
Zdziarski, Lubinski \& Smith 1999).  Specifically, increases in the UV
seed should result in a softer X-ray spectrum. Indeed the observation
of such behavior, including in NGC 7469, provides arguably the
strongest evidence for the Comptonization hypothesis (N2K; Petrucci et
al. 2000). These spectral changes can, however, have an important
effect on the flux as measured in a specific, narrow band, confusing
any correlation (N2K). In terms of the lags, Kazanas et al. (1997)
have argued that in an extended corona, covering several decades in
radius, the most rapid variations at all energy bands are dominated by
photons scattering close to the central source and corresponding to
single scatterings. In this case, we expect the short time scale
variations to exhibit only very small time lags across different
energy bands. On the other hand, the longer time scales sample photons
that have suffered multiple scatterings as well. Here the time lags
will be revealed clearly only in the cross-spectrum, as the lags
should be most prominent in the long time scale variations
(e.g. Miyamoto et al 1988, 1992).  We would, however, expect a smaller
variability amplitude at high energies due to the ``washing out'' of
the variability in the extended region.

Our detailed variability analysis broadly supports the thermal
Comptonization model just outlined. First, at least on long time
scales, we do observe a higher variability amplitude in soft X-rays
than in the hard band. This is also seen in other sources (e.g. Nandra
et al. 1997; Markowitz \& Edelson 2000).  We also confirm the
correlation of the UV seed photons with the X-ray spectrum, predicted
by Comptonization.  Cross-correlation analysis of the UV and the
softness ratio limit the time lag to be very short -- less than $\sim
8$~hours. We note and discuss later that there appear to be variations
in the ratio on time scales {\it even shorter} than variations in the
supposed seed photons, however. This may complicate the lag error
analysis and indeed as the UV does not show clear variations on time
scales as short as 8 hours, this limit may be underestimated.  It is
clear, though, that the corona responds very rapidly to changes in the
UV, implying that it is rather compact and close to the
source of the seed photons.

In addition to delays between the seed photons and the X-rays, we also
expect time delays within the Comptonized component.  Although the
intra-X-ray CCFs are peaked at zero, they are skewed towards lags in
the sense that the soft X-ray variations precede those at harder
energies. This is consistent with the idea that the harder photons
``diffusing'' through the Comptonizing medium and emerging from larger
radii. Naive fitting of the CCFs suggest that they can be deconvolved
into a narrow peak with zero lag and a broader component which is
delayed by $\sim 0.5d$. If the delay is due to light travel time
effects, it implies a very large corona (e.g. Hua et al. 1997; Kazanas
et al. 1997) corresponding to a size of $\sim 1000 M^{-1}_{7} \ R_{\rm
g}$ where $M_{7}$ is the black hole mass in units of $10^{7}$\Msun\
(cf. the estimate for NGC 7469 of $8\times 10^{6}$~\Msun; Collier et
al. 1998). Note that such a large size is not necessarily inconsistent
with the rapid variability or the peak of the CCFs at zero lag, as
this can in principle originate in the innermost parts of the corona
after single scatterings.  Another possibility is that the delays are
not due to a light travel time or the Comptonization process but, for
example, to an instability time scale in the accretion disk. The
skewing of the CCF implies an instability or flare propagating inwards
on a time scale of $\sim 1d$, first affecting the softer X-rays and
then inducing variability in the hard X-rays. In this scenario, the
sharp peaks in the CCFs should correspond to the fact that the X-ray
emitting flares respond to the UV seed variations with no delays
(i.e. an optically thin corona) as the very small lag between the UV
and the softness ratio indicates.

\subsection{Rapid variability: a coronal origin?}

Despite the broad support which our variability analysis affords the
Comptonization model, it is becoming clear that there is a variability
process distinct from the apparent Compton seed (i.e. the observed
UV). N98 had already noted that rapid variations in the 2-10 keV
X-ray flux were observed which were not present in the UV. We have
also shown that there are rapid changes in the X-ray spectrum which
occur faster than changes in the supposed seed. This implies that, if
the UV are indeed the seed photons, they cannot be solely responsible
either for the flux variability in the X-rays, or the spectral
variations via cooling.  A potential way of explaining this is that UV
acts as a surrogate for the ``true'' seed photons, the EUV, which
might show (unobserved) variations even more rapid than the
X-rays. Our observation of a ``hardening'' of the PSD as a function of
energy appears to rule this out, however.  Wherever the seed photons
originate, once they are up scattered we expect the high frequency
variability to be less at high energies.  We therefore conclude that
there must be a mechanism unrelated to the seed photons that
dominates the variability on time scales of less than about one
day. We note with great interest that the same behavior has been
observed for Cyg X-1 by Nowak et al. (1999), albeit on a time scale
$\sim 10^{6}$ times shorter. Our conclusion is the same as theirs,
that the fast variability most likely something to do with the process
which heats the corona. For example, the magnetic flare model of
Poutanen \& Fabian (1999) predicts at least that the high frequency
power spectrum should be similar at all energies.

Assuming that the X-rays are indeed produced by Compton upscattering,
it seems highly unlikely that this ``hardening'' of the PSD can be
produced in a model where the X-rays are produced in a single,
coherent region. In such a scenario, any changes in the seed photons
would affect the PSD in the opposite way to that observed. Indeed,
this hardening appears to rule out the possibility even of an
inhomogeneous, extended corona as proposed by Kazanas et
al. (1997). In this model the zero lag peaks in the CCFs are due to
contributions from single scattered photons in the central regions of
the corona, from which the rapid variability arises.  At the very
least we would then expect to see the same amplitude at high
frequencies but much more likely the variability should be drastically
reduced with the PSD rapidly {\it steepening} as a function of energy
due to the ``washing out'' effect of multiple scatterings described
above (Brainerd \& Lamb 1987; Nowak \& Vaughan 1996; Hua \& Titarchuk
1996; Kazanas \& Hua 1999).  One way out of this is by hypothesizing
that the extended corona has a temperature gradient as well as having
a density gradient, and that the hotter inner regions respond to a
very rapidly variable seed source which we do not observed (e.g. the
EUV). ADAF models, for example, exhibit such a gradient, with the
temperature being constant out to some radius, outside which it drops
as $R^{-1}$ (Narayan \& Yi 1994). It seems more likely, however, that
the data are indicating that the corona is patchy, consisting of
``flares''. Again the flares in the inner regions should typically
be hotter. They would therefore account for more of the high energy
photons, but be more rapidly variable because of their smaller size
scale. Depending on the heating mechanism, it may take time for the
electron temperature to respond to changes in the heating rate and
those changes also have a finite time scale associated with them. This
could potentially have different effects at different energies. For
example if the temperature change propagates from the inner to the
outer regions in a finite time, this would produce more rapid
variations in the hard flux in those inner regions. We await a
detailed model that explains this flattening of the PSD with
increasing photon energy.

Our interpretation is of two separate origins for the variability, one
due to variations in the seed photons and another due to the
coronal heating mechanism. The first dominates the variability on time
scales greater than about 1 day, gives a larger amplitude in the soft
X-rays and induces spectral variability by cooling the corona. The
time delay involved is very short, with the CCF of the UV vs. the
softness ratio being peaked at zero, but it is skewed to positive lags
in a similar fashion to the intra-X-ray CCFs. The second process
induces high frequency variability roughly simultaneously at all
energies, but with larger amplitude at high energies. It results in
intra-X-ray CCFs which are strongly peaked at zero, explaining the
sharp central peak in the ACFs and CCFs. Since the rapid variability
mechanism affects the harder energies more, it seems highly unlikely
that the X-ray emitter is a single region, but rather consists of
localized ``flares''. This mechanism must also induce rapid spectral
variability, because we observe changes in the X-ray hardness ratio as
short as $\sim 90$~min (1 orbit) which are more rapid than the
variations in the UV seed flux. Once again, this cannot be account for
by invoking a more rapidly-varying EUV seed because of the hardening
of the PSD. It seems reasonable for the coronal heating mechanism to
induce rapid spectral variability -- after all, the very process of
heating would results in a spectral change.

We note that a ``two-process'' variability mechanism is suggested by
some other observations, notably those of NGC 3516 (e.g. Maoz et
al. 2000; Edelson et al. 2000) which appear to show rather similar
characteristics to NGC 7469. In particular the most rapid variations
in the X-rays are not observed in the optical (Maoz et al. 2000).  The
soft X-rays vary simultaneous with the hard X-rays, but have larger
amplitude on time scales of a few days (Edelson et al. 2000).  NGC
5548 also shows this behavior, with the soft X-ray/hard X-ray CCF
having a peak at zero lag but being strongly skewed towards positive
lags (Chiang et al. 2000). Chiang et al. also show evidence for an
explicit time delay in this source, with the EUV emission leading the
X-rays by $\sim 30ks$, rather similar to the ``lag'' time scale
inferred here.  A prediction of our model is that the cross-spectrum
should show time lags at long time scales which will be smaller
on short time scales, as seen in Cyg X-1 (Miyamoto et al. 1992).

Note that, as discussed by N2K, the longer-timescale variations
induced by the seed photons need not be intrinsic to the material
producing them (e.g. the accretion disk). Energetically it is possible
that the optical/UV variations on time scales of a few days are due to
reprocessing of soft X-ray photons. This is consistent with the fact
that we observe changes in the hardness ratio faster than those in the
UV flux -- these variations being smoothed out by an extended
reprocessor, and the observation of time lags within the UV and
optical (Wanders et al. 1997; Collier et al. 1998). The reprocessed
seed variations would then cool the corona, in turn changing the
reprocessed flux and back again, in exactly the kind of ``feedback''
mechanism envisaged by Haardt \& Maraschi (1991, 1993). We favor such
a mechanism for the origin of the X-ray and variable UV radiation in
NGC 7469 and other AGN, with the added provisos that the corona is
patchy (Haardt et al. 1994; Stern et al. 1995) and the heating should
occur in rapid, localized ``bursts'' (which may amount to the same
thing). These bursts or flares cannot be static, and each must either
have a different intrinsic properties (e.g. $\tau$, $kT$, geometry) or
different evolution of those properties. Possibilities include that
mentioned above, that the inner regions are hotter and more rapidly
variable, or that the inner regions have roughly the same initial
properties, but evolve more rapidly than the inner region.

\subsection{Features in the PSD}

We observe no clear features in the PSD of NGC 7469, such as the low
frequency ``knee'' (Papadakis \& McHardy 1995; Edelson \& Nandra 1999)
or high frequency ``break'' (Nowak \& Chiang 1999). There is some
structure in the high frequency PSD which can be modeled as a ``QPO''
on a timescale of approximately 2500s. This is similar to the time
scales of QPO which have been reported from some EXOSAT observations
(Papadakis \& Lawrence 1993b; Papadakis \& Lawrence 1995).  We are
cautious about such an interpretation here, however.  The shape of the
QPO cannot be determined as it is very close to the gap between the
low-- and high--frequency PSDs and it could simply be that the high
frequency PSD is ``noisy'' and not well fit by a simple power
law. This is also seen in Cyg X-1 (Nowak et al. 1999) and has been
interpreted as there being a more complex power spectral form
(e.g. Nowak 2000). We await better data, which can be obtained with,
e.g. XMM, on the short-medium time scales before drawing firm
conclusions about NGC 7469 or other AGN.

\subsection{Rapid variability of the PSD}

A final intriguing result from the X-ray variability of NGC 7469 is
the apparent increase in the RMS variability parameter when the UV
flux is weakest and the X-ray spectrum the hardest. The behavior of
this individual object contrasts with that when comparing objects
where sources with flat spectra are {\it less} variable, at least on
$\sim d$ time scales (e.g. Green et al. 1993; Turner et al. 1999). The
reason for the behavior in NGC 7469 is not clear at this point, but it
may be related to some form of instability in the corona. For example,
as the spectrum hardens it may reach a critical point at which
electron-positron pair production becomes important, inducing
additional variability. Naively, however, we might expect this to
increase the optical depth, smoothing out rapid variability, however.
We have hypothesized above that the hotter parts of the corona occur
closer in, in order to explain the ``hardening'' of the PSD. If this
is indeed the case, then the source should be most rapidly variable
(in a normalized sense) when the spectrum is hardest, and therefore
the flux dominated by the innermost, hottest and most rapidly variable
regions.  The increase in RMS is also consistent with an increase in
the normalization of the PSD (although we are not able to rule out the
possibility that the shape changes). In the general context of shot
noise models, with flares happening randomly, the normalized PSD
amplitude can change in two ways: 1) the mean rate of flares per unit
time changes or 2) the flare shape changes in the sense that flares last
shorter or longer.  The latter seems more consistent with the
observation of an increased RMS when the flux is weak.  Whatever the
origin of these phenomena, it is clear that variability still has much
to tell us about both the radiation mechanism and the dissipation
process that causes active galaxies to emit X-rays.

\acknowledgements 
We thank the \xte\ team for their operation of the satellite,
especially Dave Smith and Keith Jahoda for their work on the
background subtraction and help producing the background light curves,
Tal Alexander and Shai Kaspi for help with the ZDCF, and Brad Peterson
for providing the code for the ICCF and lag error calculations. We
acknowledge valuable discussions with Demos Kazanas and Richard
Mushotzky. The referee, Julian Krolik, is thanked for some interesting
comments. KN is supported by NASA grant NAG5-7067 provided through the
Universities Space Research Association.

\clearpage

\begin{deluxetable}{lllll}

\tablecolumns{5}
\tablecaption{Variability parameters
\label{tab:var}}

\tablehead{
\colhead{Band} & 
\colhead{$\overline{F}$} & \colhead{$\sigma_{\rm F}$} &
\colhead{$\sigma^{2}_{\rm RMS}$ (5760s)} & 
\colhead{$\sigma^{2}_{\rm RMS}$ (16s)} 
}
\startdata
2-10     & 8.64 & 1.83 & $2.55 \pm 0.16$ & $2.78\pm 0.03$ \\
2-4      & 4.26 & 1.09 & $2.99 \pm 0.19$ & $3.26\pm 0.05$ \\
4-10     & 4.23 & 1.04 & $2.22 \pm 0.14$ & $2.44\pm 0.05$ \\
10-15    & 1.22 & 0.59 & $1.81 \pm 0.16$ & $2.43\pm 0.19$ 
%15-24    & 0.41 & 0.62 & $6.2 \pm 1.4$ & $16.9 \pm 1.7$ \\

\tablecomments{ $\overline{F}$ is the mean flux in ct s$^{-1}$;
$\sigma_{\rm F}$ is the square root of the variance of the light curve
(ct s$^{-1}$). $\sigma^{2}_{\rm RMS}$ is the excess variance in units
of $10^{-2}$ (dimensionless), calculated from the 5760s light curves
and 16s light curves.}
\enddata
\end{deluxetable}

\clearpage

\begin{deluxetable}{lllll}
\small

\tablecolumns{5}
\tablecaption{Cross-correlation results  \label{tab:ccf}}

\tablehead{
\colhead{Band} & 
\colhead{$r_{\rm 0}$ (256s)} & \colhead{Delay (ks)} & 
\colhead{$r_{\rm 0}$ (5760s)} & \colhead{Delay (ks)}
}

\startdata
2-4 keV vs. 4-10 keV       & 0.83 & $0.7 \pm 0.8$ & 0.85 & $1.4\pm 2.0$ \\
2-4 keV vs. 10-15 keV      & 0.49 & $1.1 \pm 1.9$ & 0.67 & $0.6\pm 3.0$ \\
4-10 keV vs. 10-15 keV     & 0.49 & $0.5 \pm 1.7$ & 0.73 & $0.0\pm 2.3$ \\
UV vs. 2-4/4-10 keV        & \nodata & \nodata & 0.44 & $-8.1\pm 30.3$ 
%2-4/10-15 keV  & \nodata & \nodata & 0.42 & $16.7\pm 66.7$ \\
%4-10/10-15 keV & \nodata & \nodata & 0.18 & \nodata \\

\tablecomments{
$r_{\rm 0}$ is the correlation coefficient at zero lag. As the zero
lag points have been suppressed in the X-ray/X-ray correlations, due to
correlated errors, this was estimated as the average of the two
adjacent bins; the delays were calculated using the method of Peterson
et al. (1998).  
} 
\enddata
\end{deluxetable}

\clearpage

\begin{deluxetable}{lccccc}
\tablecolumns{6}
\tablecaption{Results of Power Law model fitting to the  
Low and High frequency PSD of NGC~7469
\label{tab:pds_sep}
}

\tablehead{
\colhead{Band} & 
\colhead{$\alpha_{low}$} & 
\colhead{A$_{low}$ (Hz$^{-1})$} & 
\colhead{$\alpha_{high}$} & 
\colhead{A$_{high}$ (Hz$^{-1})$} & 
\colhead{$\chi^{2}$/dof} 
}
\startdata
 $2-10$ keV     & $1.58^{+0.16}_{-0.15}$ & $3.8^{+1.1}_{-0.9}$ & 
$1.72^{+0.39}_{-0.32}$ & $5.8^{+6.2}_{-2.8}$ & 555.1/572 \\
 $2-4$ keV      & $1.58^{+0.16}_{-0.17}$ & $4.5^{+1.7}_{-1.1}$ & 
$1.29^{+0.72}_{-0.35}$ & $2.7^{+4.4}_{-1.6}$ & 607.4/572 \\
 $4-10$ keV      & $1.57^{+0.18}_{-0.17}$ & $3.3^{+1.2}_{-0.9}$ & 
$1.99^{+0.70}_{-0.48}$ & $12.7^{+25.4}_{-7.4}$ & 588.7/572 \\
 $10-15$ keV      & $1.48^{+0.23}_{-0.22}$ & $3.9^{+2.6}_{-1.8}$ & 
$1.94^{+0.65}_{-0.48}$ & $62.9^{+105.7}_{-37.7}$ & 560.3/572 
% $15-24$ keV      & $0.64^{+0.30}_{-0.30}$ & $73.4^{+55.8}_{-38.2}$ &
%$2.80^{+1.20}_{-0.85}$ & $2880^{+7660}_{-1670}$ & 591.4/572 \\

\tablecomments{
$A$ is the normalization of the PSD at $10^{-4}$~Hz and
$\alpha$ the slope. Parameters marked ``low'' refer to the low frequency PSD
below $\sim 10^{-4}$ Hz and ``high'' refer to the high frequency PSD
above $\sim 10^{-3.5}$ Hz}

\enddata
\end{deluxetable}

\clearpage

\begin{deluxetable}{lccc}
\tablecolumns{4}
\tablecaption{Results of the Power Law model fitting to the combined
Low+High frequency PSD of NGC~7469 
\label{tab:pds_joint}
} 
\tablehead{
\colhead{Band} & 
\colhead{$\alpha$} & 
\colhead{$A$ (Hz$^{-1})$} & 
\colhead{$\chi^{2}$/dof} 
}

\startdata
 $2-10$ keV	& $1.55^{+0.10}_{-0.06}$ & $4.0^{+0.6}_{-0.6}$ & 591.5/574 \\
 $2-4$ keV	& $1.57^{+0.10}_{-0.08}$ & $4.7^{+0.8}_{-0.8}$ & 608.2/574 \\
 $4-10$ keV	& $1.46^{+0.07}_{-0.08}$ & $4.2^{+0.7}_{-0.6}$ & 593.6/574 \\
 $10-15$ keV	& $1.12^{+0.10}_{-0.08}$ & $9.4^{+1.9}_{-1.9}$ & 579.2/574
% $15-24$ keV 	& $0.80^{+0.08}_{-0.08}$ & $63^{+18}_{-13}$ & 619.1/574

\enddata
\end{deluxetable}

%-----------------------------------------------------------------------
%\newpage

\clearpage

\appendix

\section{Effects of residual background variations on the power spectra}

As the ACF, CCF and PSD techniques is rather sensitive, particularly
to periodic signals, our results could be strongly affected by any
residual uncertainty in the background subtraction. We now investigate
such affects using both a 4d observation of a ``blank sky'' pointing
and the high energy data for NGC 7469, which are more dominated by
background.

\subsection{The 4d background}

%\begin{figure}
%\epsscale{0.8}
%\plotone{./figs/fig_bgd_hf.ps}
%\caption{High frequency periodogram of the 4 day background observation
%in (top to bottom) 2-4, 4-10, 10-15 and 15-24 keV bands. The dashed
%line shows the 95\% confidence limit assuming the periodogram consists
%only of the expected Poisson noise, with no signal. Only one
%peak is seen above this level, in the 10-15 keV periodogram, and this
%is at a frequency deep within the noise of the NGC 7469 PSD. 
%\label{fig:bgd_hf}}
%\end{figure}

\begin{figure}
\epsscale{0.8}
\plottwo{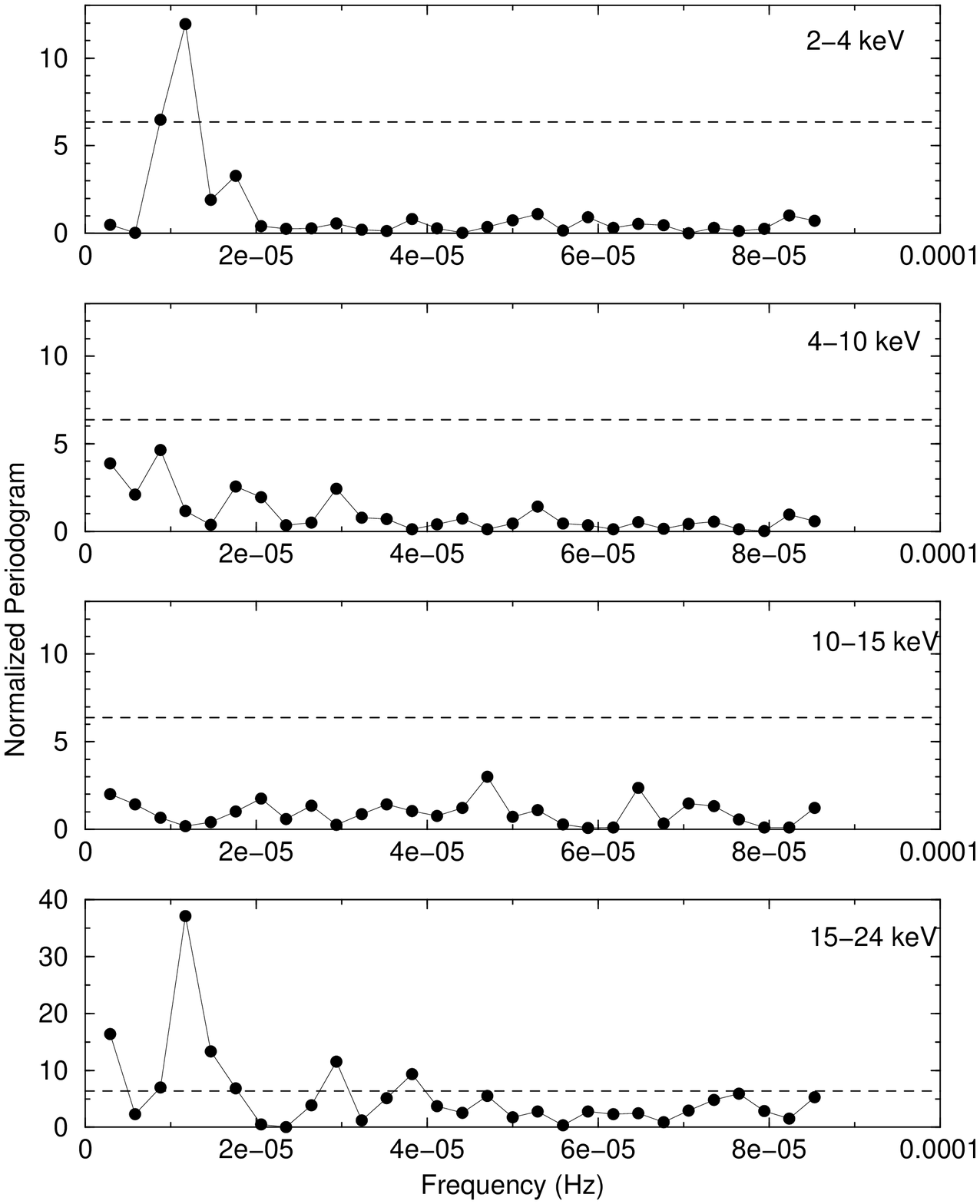}{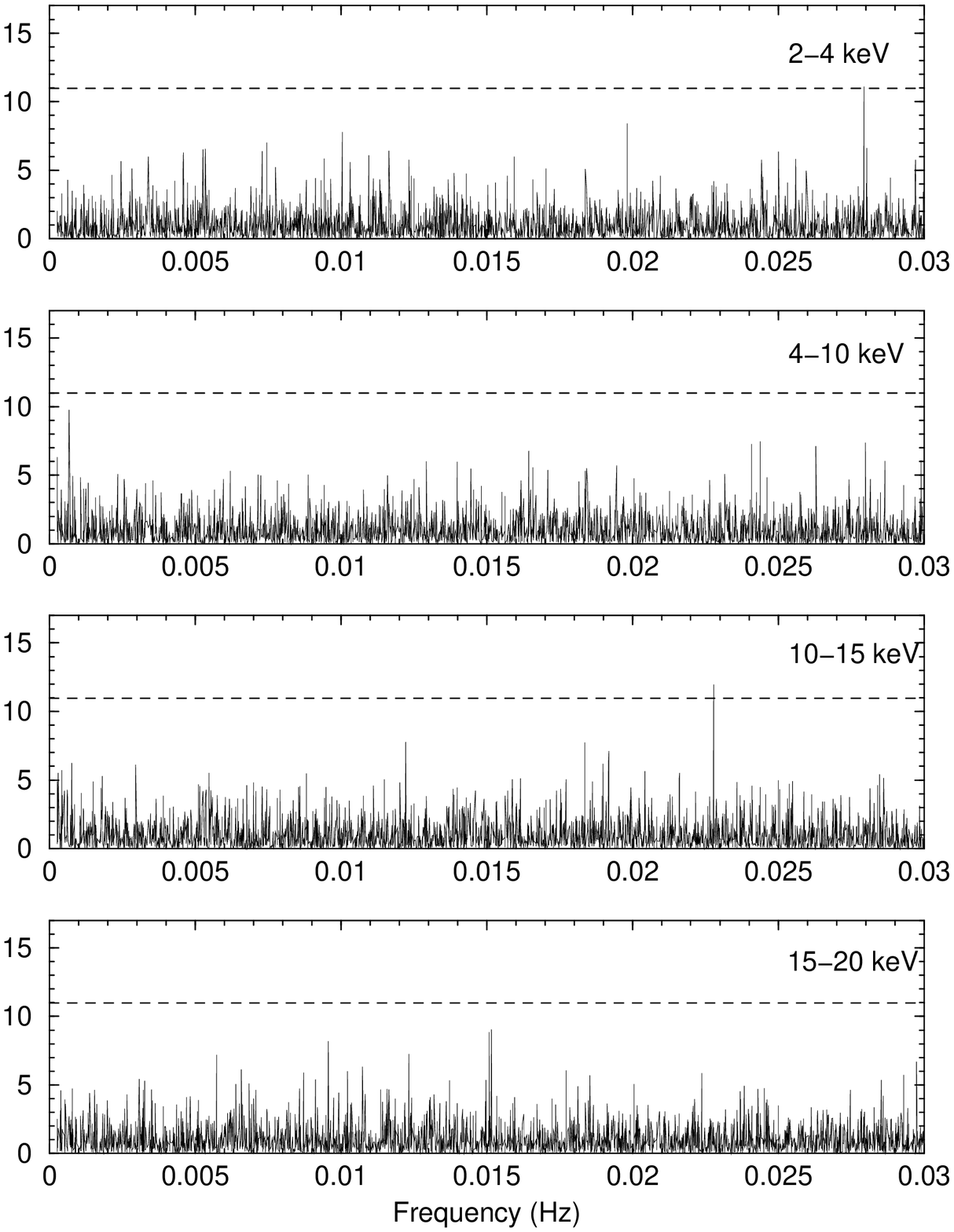}
\caption{(left panels) Low frequency periodograms of the 4 day
background observation in (top to bottom) 2-4, 4-10, 10-15 and 15-24
keV bands.  The dashed line shows the 95\% confidence limit assuming
the periodogram consists only of the expected Poisson noise, with no
signal.  A significant signal is observed on a time scale of $\sim 1d$
in the 15-24 keV light curve, which is also visible but much less
prominent in the 2-4 keV light curve. (right panels) high frequency
periodograms in the same bands.  Only one peak is seen above this
level, in the 10-15 keV periodogram, and this is at a frequency deep
within the noise of the NGC 7469 PSD.
\label{fig:bgd_period}}
\end{figure}

\begin{figure}
\epsscale{0.8}
\plotone{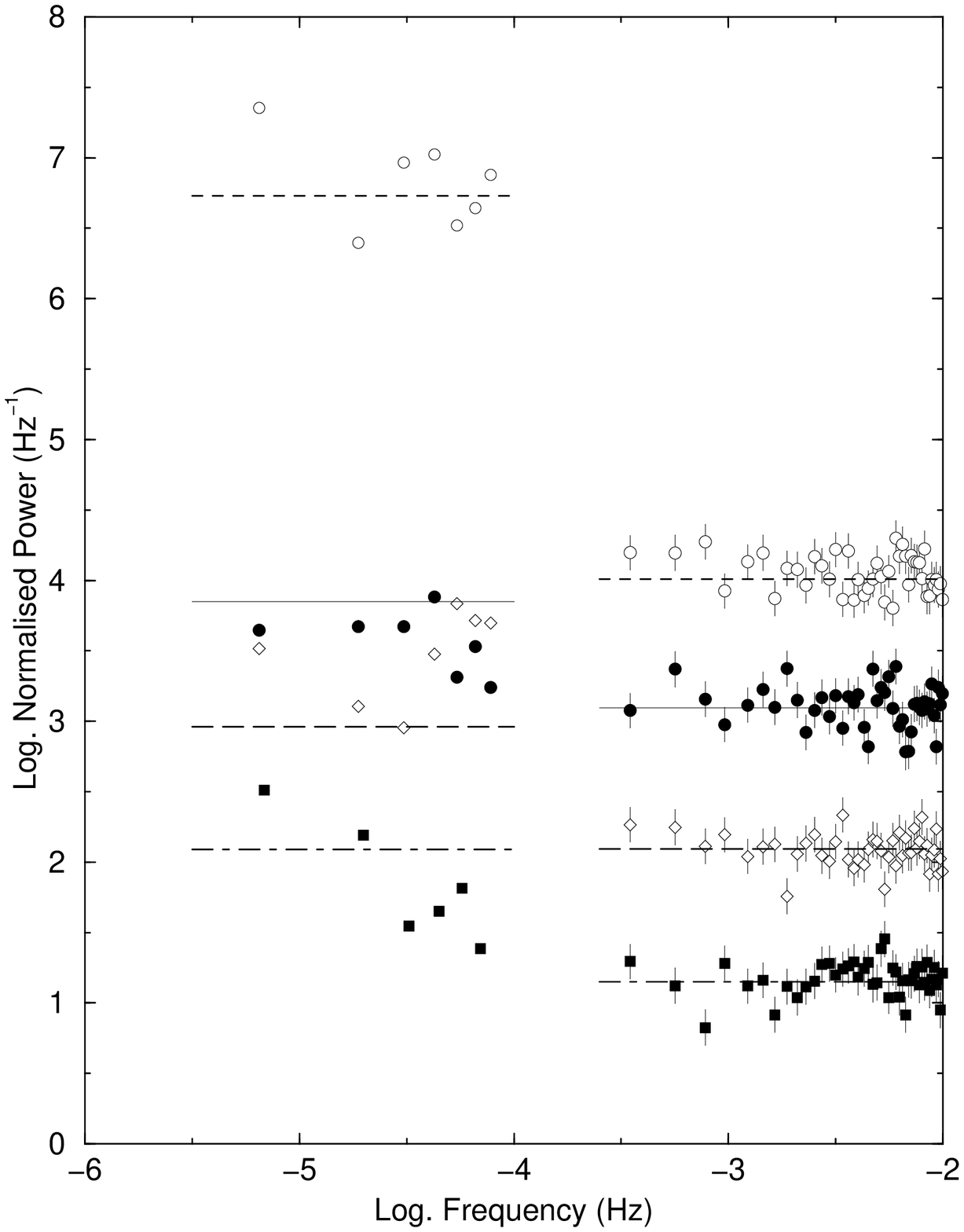}
\caption{PSD of the 4 day background observation in 4 bands. 
Open circles are the 15-24 keV power spectrum,
filled circles are the 10-15 keV, open diamonds the 4-10 keV, and
filled squares the 2-4 keV. Short dashed line, solid line, long-dashed
line and dotted-dashed line are the expected Poisson noise level
in each band. 
\label{fig:bgd_pds}}
\end{figure}

As part of the development of the background model, \xte\ makes
observations of several points in the sky which are free from bright
sources - so-called ``blank sky'' pointing. One such observation was
made near-continuously for $\sim 4$d, which was intended to identify
medium-timescale trends and periodicities in the background. Though
shorter, this dataset is the blank sky pointing which most closely
resembles our NGC 7469 dataset, and can be used to search for
systematic effects which might bias our results.  We applied the
background subtraction model to this dataset in the same way as for
NGC 7469, and extracted light curves (using a bin of size 16 and 5760
sec) in the same four energy bands. We used these light curves and the
same method as with the NGC~7469 data to estimate the power spectrum
at high and low frequencies respectively.  The unbinned periodograms
are shown in Fig.~\ref{fig:bgd_period}.

Under the hypothesis that the background model describes the observed
background properly their power spectrum should be constant and equal
to the constant Poisson noise power level ($\Delta t\times
\sigma^{2}$, where $\sigma$ is the variance of the light curve). Note
that the means of the background-subtracted light curves are not zero,
with the 2-4, 4-10, 10-15 and 15-24 keV light curves showing offsets
from zero relative to the predicted background of 3.5, 3.7, 1.5 and
-0.2 per cent levels respectively. These contributions, and the
reduction in them with increasing energy, is consistent with a
positive diffuse background fluctuation in the direction of the
background pointing, which should be constant and will not affect the
power spectra. When we subtract off the mean value from the light
curves a residual constant component will be removed. We note that
this may have a small effect on our calculation of the RMS values, due
to a small inaccuracy in the calculation of the mean, but this is
smaller than the associated statistical error (at least for the 5760s
light curves) and so can be ignored. The periodograms in
Fig.~\ref{fig:bgd_period} have been divided by the expected Poisson
power level, in which case the points should have a \chisq\
distribution with two degrees of freedom (Papadakis \& Lawrence
1993a). Normalized in this way, we can calculate a confidence level
from the periodogram to exceed a certain value, and the 95 per cent
confidence limit (accounting for the number of points) is shown as the
dashed line in the Figure. Examining first the high frequency
periodogram (right panels of Fig.~\ref{fig:bgd_period}) we see that,
although there are a few spikes only one exceeds the 95 per cent
confidence level. This is in the 10-15 keV light curve at a frequency
$\sim 2.3 \times 10^{-2}$, well into the noise of the NGC 7469 PSD
(c.f. Fig.~\ref{fig:pds}). We conclude that residual errors in the
background subtraction will not affect the high frequency PSD. As for
the low frequencies (left panels of Fig.~\ref{fig:bgd_period}), the
15-24 keV periodogram shows a sharp peak at $\sim 10^{-5}$s, which is
also present, but at a much lower level, in the 2-4 keV band. This
probably represents a daily periodicity, which we discuss further
below.

\begin{figure}
\epsscale{1.0}
\plotone{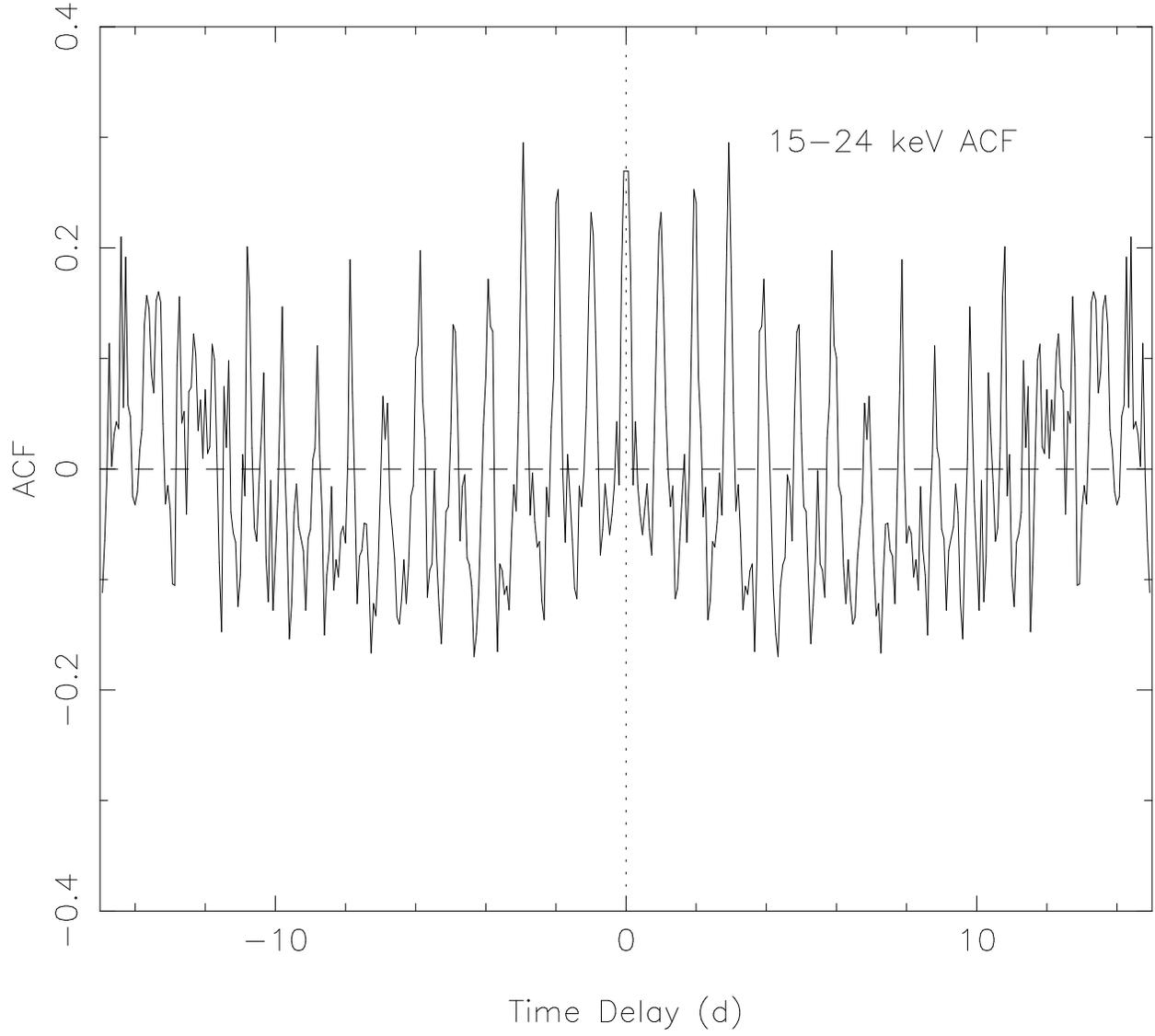}
\caption{ACF of the 15-24 keV light curve of NGC 7469, calculated
by the interpolation method. This shows peaks every day which are
almost certainly due to residual background. 
\label{fig:vh_acf}}
\end{figure}

Fig.~\ref{fig:bgd_pds} shows the combined background power spectra,
this time normalized to the mean squared as the NGC~7469 data, in log
space and with the high frequency PSD binned like that of NGC 7469.
They are flat and consistent with the expected power level.  We find
no systematic excess of power at high frequencies for the hard X-ray
light curve, which might account for the ``hardening'' of the PSD just
noted. Neither do we find a clear feature at $\times 10^{-3.5}$~Hz
where an excess is observed in the NGC 7469 PSD.  Furthermore we find
no systematic differences between the four power spectra.  Although we
note that this 4d PSD is not as high quality as the 30d PSDs we
derived for NGC 7469, this analysis indicated that our conclusion for
NGC 7469 are robust and secure.

\subsection{The high energy NGC 7469 data}

The PCA team (Jahoda, priv. comm) have presented evidence for a daily
periodicity like that seen in Fig.~\ref{fig:bgd_period} in the 30-70 keV
light curves of various datasets, attributed to residual errors in the
background. In our analysis of NGC 7469, we had initially hoped to
extend our analysis above 15 keV, where the effects of PSD hardening
would have been much more pronounced. We found, however, that the 
15-24 keV bands was strongly affected by this daily periodicity.

\begin{figure}
\epsscale{0.8}
\plotone{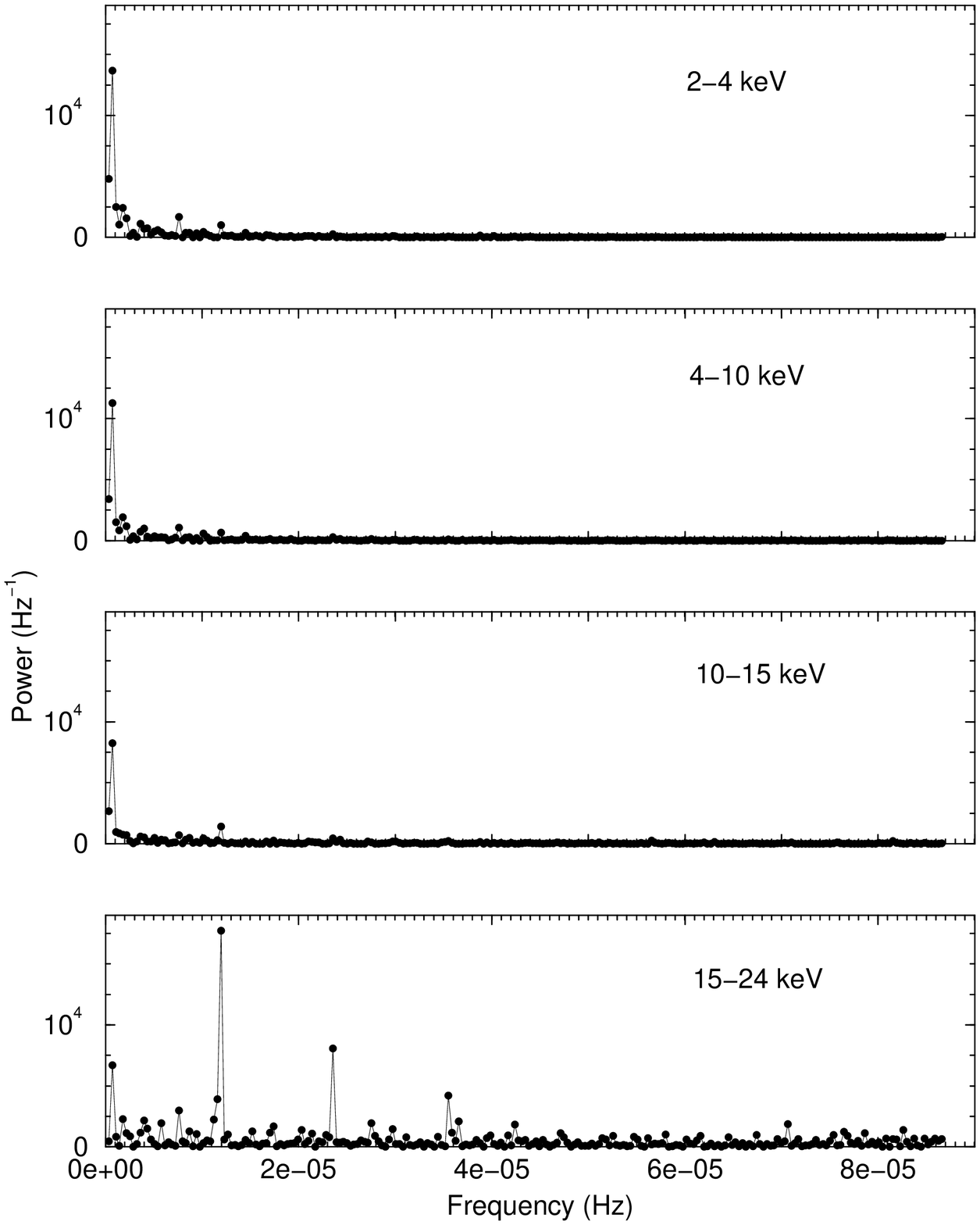}
\caption{Unbinned low frequency PSD of NGC 7469 in four energy bands
(top to bottom): 2-4 keV, 4-10 keV, 10-15 keV and 15-24 keV, plotted
in linear space. The 15-24 keV PSD shows a very clear periodicity on a 
timescale of 1 day, and associated harmonics. This periodicity is due to
daily variations in the background which have not been modeled
correctly. The periodicity is very much weaker in the other bands, and
has a negligible effect on them. 
\label{fig:lin_pds}}
\end{figure}

Fig.~\ref{fig:vh_acf} shows the autocorrelation function of the NGC
7469 light curve in the 15-24 keV band. This clearly shows very
strong, daily peaks. There is very little signal in the ACF other than
this daily variations, which show that they dominate the variability
in this band. Note that the ACFs in the lower energy bands do not show
a strong signal on this time scale, with the source signal dominating
(Fig.~\ref{fig:acf}). Nonetheless, even if present at a lower level in
the in the other energy bands they could bias our estimate of the PSD
and introduce spurious signals in the CCFs. We do not believe this is
a strong effect, however. Fig.~\ref{fig:lin_pds} shows the unbinned,
low frequency periodogram in linear space for the four bands between 2
and 25 keV.  As expected, the 15-24 keV plot shows a very strong peak
at the daily time scale, and also strong harmonics. The other
periodograms also show peaks but these are much smaller and will not
significantly affect our estimates of the power spectra. We have
tested this by removing the data points at the daily periodicity and
the two harmonics from the lower energy periodograms, binning them and
comparing to the binned PSDs in Fig.~\ref{fig:3band_pds}. No
significant difference is observed, leaving us confident that
background effects do not bias our results. We particularly note that
there are no significant biases as a function of energy that would indicate
problems with our conclusion of the PSD hardening.


\clearpage
\begin{references}

\reference{ab91} Abramowicz, M.A., Bao, G., Lanza, A., Zhang, X.-H., 1991, 
                 A\&A, 245, 454
\reference{al97} Alexander, T., 1997, in ``Astronomical Time 
                 Series'', D. Maoz, A. Sternberg, E. Leibowitz, Eds, 
                 Kluwer, Dordrecht, p. 163
\reference{ba96} Bao, G., Abramowicz, M.A., 1996, ApJ, 465, 646
\reference{bh90} Belloni, T., Hasinger, G., 1990, A\&A, 230, 103
\reference{bl77} Brainerd, J., Lamb, F.K., 1987, ApJ, 317, L33
\reference{ch00} Chiang, J., et al., 2000, ApJ, 528, 292
\reference{cl92} Clavel, J., \etal, 1992, ApJ, 393, 113
\reference{co98} Collier, S., et al., 1998, ApJ, 500, 162
\reference{do90} Done, C., Ward, M.J., Fabian, A.C., Kunieda, H.,
                 Tsuruta, S., Lawrence, A., Smith, M.G., Wamsteker, W.,
	         1990, MNRAS, 243, 713
\reference{ed00} Edelson, R.A., \etal, 2000, ApJ, 534, 180
\reference{ek88} Edelson, R.A., Krolik, J.H., 1987, ApJ, 333, 646
\reference{en99} Edelson, R.A., Nandra, K., 1999, ApJ, 514, 682
\reference{ed96} Edelson, R.A., \etal, 1996, ApJ, 470, 364
\reference{gp87} Gaskell, C.M., Peterson, B.M., 1986, ApJS, 65, 1
\reference{go96} Gondek, D., Zdziarski, A.A., Johnson, W.N., George, I.M.,
                 McNaron-Brown, K., Magdziarz, P., Smith, D., Gruber, D.E.,
                 1996, MNRAS, 282, 646
\reference{gr93} Green, A.R., McHardy, I.M., Lehto, H.J., 1993, MNRAS, 
                 265, 664
\reference{hm91} Haardt, F., Maraschi, L., 1991, ApJ, 380, 51
\reference{hm93} Haardt, F., Maraschi, L., 1993, ApJ, 413, 507
\reference{ha94} Haardt, F., Maraschi, L., Ghisellini, G., 1994, ApJ, 432, L95
\reference{ha97} Haardt, F., Maraschi, L., Ghisellini, G., 1997, ApJ, 476, 620
\reference{ht96} Hua, X.-M., Titarchuk, L., 1996, ApJ, 469, 280
\reference{hu97} Hua, X.-M., Kazanas, D., Titarchuk, L., 1997, ApJ, 482, L57
\reference{ka97} Kazanas, D., Hua, X.-M., Titarchuk, L., 1997, ApJ, 480, 735
\reference{ka97} Kazanas, D., Hua, X.-M., 1999, ApJ, 519, 750
\reference{ja96} Jahoda, K., Swank, J.H., Giles, A.B., Stark, M.J., 
                 Strohmayer, T., Zhang, W., Morgan, E.H., 1996,
                 EUV, X-ray and Gamma-ray Instrumentation for Space Astronomy 
                 VII, O. H. W. Siegmund and M. A. Grummin, Eds.,
                 SPIE 2808, p. 59
\reference{kr00} Kriss, G.A., Peterson, B.M., Crenshaw, D.M, Zheng, W., 2000, 
		 ApJ, 535, 58
\reference{lp93} Lawrence, A., Papadakis, I., 1993, ApJ, 414, L85
\reference{la85} Lawrence, A., Watson, M.G., Pounds, K.A., Elvis, M., 
                   1985, \mnras, 217, 685
\reference{le96} Leighly, K., Mushotzky, R.F., Yaqoob, T., Kunieda, H.,
		 Edelson, R.A., 1996, ApJ, 469, 147
\reference{ma82} Malkan, M.A., Sargent, W.L., 1982, \apj, 254, 22
\reference{ma00} Maoz, D., Edelson, R.A., Nandra, K., 2000, AJ, 119, 119
\reference{me01} Markowitz, A., Edelson, R.A., 2001, ApJ, in press
\reference{ma97} Marshall, H., et al., 1997, ApJ, 479, 222 
\reference{mh87} McHardy, I.M., Czerny, B., 1987, \nat, 325 696
\reference{mh98} McHardy, I.M., Papadakis, I., Uttley, P., 1998, in 
		 The Active X-ray Sky: results from BeppoSAX and RXTE, 
		 L. Scarsi, H. Bradt, P. Giommi, and F. Fiore, Eds. 
		 (Amsterdam: Elsevier), p. 509
\reference{mi92} Miyamoto, S., Kitamoto, S., Mitsuda, K., Dotani, T., 1988,
		 Nature, 336, 450
\reference{mi92} Miyamoto, S., Kitamoto, S., Iga, S., Negoro, H., Terada, K.,
		 1992, ApJ, 391, 21L
\reference{mo92} Molendi, S., Maraschi, L., Stella, L., 1992, MNRAS, 255, 27 
\reference{ny94} Narayan, R., Yi, I., 1994, ApJ, 428, L13
\reference{nm97} Nayakshin, S., Melia, F., 1997, ApJ, 490, L13
\reference{na98} Nandra, K., et al.,  1998, ApJ, 505, 594 (N98)
\reference{na00} Nandra, K., et al., 2000, ApJ, 544, 734 (N2K)
\reference{na97} Nandra, K., George, I.M, Mushotzky, R.F., Turner, T.J.,
        Yaqoob, T., 1997, ApJ, 476, 70
\reference{na94} Nandra, K., Pounds, K.A., 1994, \mnras, 268, 405
\reference{no00} Nowak, M., 2000, MNRAS, 318, 361
\reference{nc00} Nowak, M., Chiang, J., 2000, ApJ, 531, L13
\reference{no99} Nowak, M.A., Vaughan, B.A., Wilms, J., Dove, J. B., 
		 Begelman, M. C., 1999, ApJ, 510, 874
\reference{pl93a} Papadakis, I., Lawrence, A., 1993a, MNRAS, 261, 612
\reference{pl93b} Papadakis, I., Lawrence, A., 1993b, Nature, 361, 233
\reference{pl95} Papadakis, I., Lawrence, A., 1995, MNRAS, 272, 161
\reference{pm95} Papadakis, I., McHardy, I.M.,  1995, MNRAS, 273, 923
\reference{pe98} Peterson, B.M., Wanders, I., Horne, K., Collier, S.,
                 Alexander, T., Kaspi, S., Maoz, D., 1998, PASP, 110, 660
\reference{pe00} Petrucci, P.-O. et al., 2000, ApJ, 540, 131
\reference{pk95} Pietrini, P., Krolik, J.H., 1995, ApJ, 447, 526
\reference{pf99} Poutanen, J., Fabian, A.C., 1999, MNRAS, 306, L31
\reference{sh76} Shapiro, S.L., Lightman, A.P., Eardley, D.M., 1976,
		 ApJ, 204, 187
\reference{st95} Stern, B.E., Poutanen, J., Svensson, R., 
                 Sikora, M., Begelman, M.C., 1995, ApJ, 449, L13
\reference{st80} Sunyaev, R.A., Titarchuk, L.G., 1980, A\&A, 86, 121
\reference{tu99} Turner, T.J., George, I.M., Nandra, K., Turcan, D., 1999, 
		 ApJ, 524, 667
\reference{ut00} Uttley, P., McHardy, I.M., Papadakis, I.E., Cagnoni, I.,
		 Fruscione, A., 2000, MNRAS, 312, 880
\reference{wa97} Wanders, I., \etal, 1997, ApJS, 113, 69
\reference{we99} Welsh, W.F., 1999, PASP, 111, 1347
\reference{wp94} White, R.J., Peterson, B.M., 1994, PASP, 106, 879
\reference{wi91} Wiita, P.J., Miller, H.R., Carini, M.T., Rosen, A., 1991,
		 in Structure and Emission Properties of Accretion Disks, 
		 ed. C. Bertout et al. (Gif-sur-Yvette: Ed. Frontieres), p557
\reference{zd94} Zdziarski, A.A., Fabian, A.C., Nandra, K., Celotti, A.,
                 Rees, M.J., Done, C., Coppi, P.S., Madejski, G.M., 1994,
                 MNRAS, 269, L55
\reference{zd99} Zdziarski, A.A., Lubinski, P., Smith, D.A., 1999, 
		 MNRAS, 303, L11


\end{references}
\end{document}